\documentclass[prx,aps,twocolumn,superscriptaddress,floats,floatfix]{revtex4}
\usepackage{amssymb,amsmath}
\usepackage[pdftex]{graphicx}
\usepackage[usenames,dvipsnames]{color}
\usepackage{soul}
\usepackage{subfigure}
\usepackage{epsfig}
\begin{document}
\title{The statistical properties of spiral- and scroll-wave turbulence in cardiac tissue}
\author{K.V. Rajany}
\email{rajany@physics. iisc. ernet. in}
\affiliation{Centre for Condensed Matter Theory, Department of Physics, Indian
Institute of Science, Bangalore 560012, India.} 
\author{Anupam Gupta}
\affiliation{Laboratoire de G\'{e}nie Chimique, Universit\'{e} de Toulouse, 
ENSIACET, INPT-UPS, 31030, Toulouse, France.}
\author{Alexander V. Panfilov}
\affiliation{Department of Physics and Astronomy, Gent University, Krijgslaan 281, S9, Gent 9000, Belgium}
\altaffiliation[\\ also at~]{Moscow Institute of Physics and Technology (State University),
Dolgoprudny, Moscow Region, Russia.}
\author{Rahul Pandit}
\email{rahul@physics. iisc. ernet. in}
\altaffiliation[\\ also at~]{Jawaharlal Nehru Centre For Advanced
Scientific Research, Jakkur, Bangalore, India.}
\affiliation{Centre for Condensed Matter Theory, Department of Physics, Indian
Institute of Science, Bangalore 560012, India.} 
\date{\today}
\begin{abstract}

Disorganized electrical activity in the heart, which is often referred to as
electrical turbulence, leads to sudden cardiac death. However, to what extent 
can  this
electrical turbulence be viewed as  classical, fluid turbulence, which is considered as a central problem of great importance in modern physics? In this paper we examine, for the first time, the statistical properties of electrical turbulence in two- and
three-dimensional generic models of cardiac tissue by using approaches employed in studies of classical turbulence. In particular, we investigate, via extensive direct numerical simulations, the statistical properties of spiral- and scroll-wave turbulence in two- and three-dimensional excitable media by using the Panfilov and the Aliev-Panfilov mathematical
models for cardiac tissue. We use very large simulation domains, and  perform 
state-of-the-art simulations on graphics processing units (GPUs), so that we
can compare the statistical properties of spiral- and scroll-wave electrical turbulence
with the statistical properties of  homogeneous and isotropic
two- and three-dimensional fluid turbulence. We show that, once electrical-wave
turbulence has been initiated,  there is a forward cascade,  in which spirals
or scrolls  form, interact, and break to yield a turbulent state that is
statistically steady and, far away from boundaries, is statistically
homogeneous and isotropic. For the transmembrane potential $V$ and the slow
recovery variable $g$, which define our models for cardiac tissue, we define
$E_V(k)$ and $E_g(k)$, the electrical-wave, spectral analogs of the fluid
energy spectrum $E(k)$ in fluid turbulence; and we show that $E_V(k)$ and
$E_g(k)$ are spread out over several decades in $k$. Thus, as in fluid
turbulence, spiral- and scroll-wave turbulence involves a wide range of spatial
scales. Furthermore, $E_V(k)$ and $E_g(k)$ show approximate power laws, in some
range of $k$; unlike fluid-turbulence, the exponents for these power laws cannot, so far, 
{{{be determined as accurately as their fluid-turbulence counterparts.}}} There are diffusive terms in
the equations we consider, but they do not dissipate spiral or scroll waves
completely because of the excitability of the medium, so no external forcing is
required to maintain these states of spiral- or scroll-wave turbulence (this is
unlike fluid turbulence that requires external forcing to reach a statistically
steady state). We show that for spiral- or scroll-wave turbulence the dimensionless ratio $L/\lambda$
is a convenient control parameter like the Reynolds number for fluid
turbulence, where $L$ is the linear size of the simulation domain and $\lambda$ the wavelength of a plane wave in the excitable medium. We calculate several other statistical properties for spiral- and
scroll-wave turbulence and, by comparing them with their fluid-turbulence
counterparts, we show that, although spiral- and scroll-wave turbulence have
some statistical properties like those of fluid turbulence, overall these types
of turbulence are special and differ in important ways from fluid turbulence.

\end{abstract}

%
%
%\pacs{}

\maketitle
\graphicspath{{./2dfigures/}{./3dfigures/}}
%%%%%%%%%%%%%%%%%%%%%%%%%
\section{Introduction}

Heart disease remains the number-one, global cause of death, with 17.3 million
deaths each year, according to the latest report of the American Heart
Association~\cite{mozaffarian}. In many cases this arises because of
\textit{sudden cardiac death}, which is a result of a cardiac arrhythmia called
ventricular fibrillation (VF). Ventricular fibrillation is a result of 
complicated, spatiotemporal dynamics of nonlinear waves of electrical
excitation in the heart. This dynamics is associated with multiple vortices
called \textit{spiral waves}, in two dimensions, and \textit{scroll waves}, in
three dimensions.  Spiral waves are found in many types of excitable media,
e.g., in Belousov-Zhabotinsky-type chemical reactions~\cite{BZ}, in the
oxidation of carbon monoxide on the surface of
platinum~\cite{bar,pande,imbihl}, in the propagation of calcium-ion waves in
Xenopus oocytes~\cite{clapham}, in the aggregation of dictyostelium discoideum
by cyclic-AMP signalling~\cite{tysonmurray,weijer} and, most important, in the
propagation of waves of electrical activation in cardiac
tissue~\cite{mines,davidenko}.

Spiral or scroll waves in cardiac tissue tend to break down into complex
spatiotemporal structures, which lead to VF, often referred as electrical
turbulence~\cite{winfree}. It is important to study the extent to which this
electrical turbulence is similar to  classical, fluid turbulence, which is a problem
of central importance in physics, engineering, and mathematics. Are the methods
developed for studies of fluid turbulence applicable to electrical-wave
turbulence in cardiac tissue?  Unfortunately, we cannot answer these questions
based on direct experimental and clinical data as most experimental studies of
turbulence require data from various space or time scales, which can differ by
orders of magnitude. Fortunately, we can use alternative approaches, such as
mathematical modelling and computational studies; indeed, these provide the
most efficient ways for studying spiral and scroll waves and, thus, cardiac
arrhythmias,  in controlled \textit{in silico} conditions~\cite{noble}.

Therefore, we examine, for the first time, the statistical properties of
electrical turbulence in two- and three-dimensional generic models of cardiac
tissue by initiating extensive, direct numerical simulations (DNSs) of spiral-
and scroll-wave turbulence in very large simulation domains, in both two
dimensions (2D) and three dimensions (3D), for two generic
models~\cite{Panfilov-model,Aliev-Panfilov-model} for cardiac tissue; these
model are used widely in cardiac research. We characterize such turbulence by
using measures that are commonly used in the statistical characterization of
fluid turbulence; these measures include Fourier-space spectra, probability
distribution functions (PDFs) and structure functions~\cite{frisch}. 

From a general point of view we can have  the following two types  of  homogeneous and
isotropic, fully developed, 3D fluid turbulence: (A) We can have  unforced, decaying turbulence, in which the energy is injected initially at
large length scales $L_I$, i.e., into a few, low-wave-number modes ($\sim
2\pi/L_I$); this energy \textit{cascades down to small length scales}, \`{a} la
Richardson~\cite{frisch}, until it reaches the dissipation scale $\eta_d$ at
which viscous losses become significant; and then the turbulence decays slowly;
in this period of slow decay, fluid turbulence displays a power-law energy
spectrum $E(k)\sim k^{-\alpha}$ for a long time and for the inertial range of
scales $2\pi/L_I \ll k \ll 2\pi/\eta_d$; at the simplest level, the
phenomenological theory~\cite{frisch,K41} of Kolmogorov (henceforth K41) yields
$\alpha^{K41} = 5/3$, a universal exponent. (B) We can also have a statistically steady turbulence, which  is an example of a driven system with a
nonequillibrium statistically steady state in which the energy input, typically
at large length scales $L_I$, is balanced by energy dissipation, which becomes
significant at length scales smaller than $\eta_d$; and, in the statistically
steady state and, in the inertial range, $E(k) \sim k^{-\alpha}$, with
$\alpha^{K41} = 5/3$.  In addition, we discuss below, various other probability
distribution functions (PDFs) and structure functions (e.g., moments of the PDF
of the differences of the fluid velocity at two points) that are used to
characterize this turbulent state. Furthermore, 2D fluid turbulence is
qualitatively different from its 3D counterpart (see below); in particular, it
displays an \textit{inverse cascade} of energy from the injection scale $L_I$
to large lengths and a \textit{forward cascade} of enstrophy (the mean-square
vorticity) from $L_I$ to smaller length scales; and $E(k) \sim k^{-\alpha}$,
with $\alpha = 5/3$ and $\alpha = 3$ in inverse- and forward-cascade regions,
respectively.

Our goal is to explore excitable-media, spiral- and scroll-wave-turbulence
analogs of 2D and 3D statistically homogeneous and isotropic fluid turbulence.
Note that there is an important qualitative difference between spiral- and
scroll-wave turbulence in excitable media and fluid turbulence insofar as
turbulence in an excitable medium is neither decaying nor forced. Once
turbulence has been initiated in an excitable medium, e.g., by using an initial
condition with a single spiral or scroll wave, we show that there is a forward
cascade, which yields small spirals or scrolls; these form, interact and break
all the time. The resulting turbulent state is statistically steady and, far
away from boundaries, is statistically homogeneous and isotropic. For the
fields $V$ and $g$, we define $E_V(k)$ and $E_g(k)$, the spectral analogs of
$E(k)$ in fluid turbulence; and we show that these spectra are spread out over
several decades in $k$, i.e., like fluid turbulence, spiral- and scroll-wave
turbulence involves a wide range of spatial scales; $E_V(k)$ and $E_g(k)$ show
approximate power laws in some range of $k$, but the exponents of these do not
appear to be universal. Even though there are diffusive terms in the equations
we consider, they do not dissipate spiral or scroll waves completely because of
the excitability of the medium. No external forcing is required to maintain
these states of spiral- or scroll-wave turbulence. The only requirements are
(i) a suitable initial condition and (ii) $L/\lambda \to \infty$ (or
$L/\lambda$ large enough in a practical calculation), where $L$ is the linear
size of the simulation domain and $\lambda$ the wavelength of a plane wave in
the excitable medium. The dimensionless ratio $L/\lambda$ is a convenient,
Reynolds-number-type control parameter ($L$ is also used as a control parameter
in a suitably scaled version of the Kuramoto-Sivashinsky
equation~\cite{jay93}). 

The remaining part of this paper is organized as follows. In Sec. II we
introduce the Panfilov~\cite{Panfilov-model} and the
Aliev-Panfilov~\cite{Aliev-Panfilov-model} mathematical models for cardiac
tissue. We then describe the numerical methods we use to study them and the
statistical measures we employ to characterize the statistical properties of
turbulence in these models. Section III is devoted to our results. Section IV
contains our conclusions and a discussion of our results.

%%%%%%%%%%%%%%%%%%%%%%%%%%%%%%%%%%%%%%%%%%%%%%
\begin{figure*}
\begin{center}
\includegraphics[width=1\linewidth]{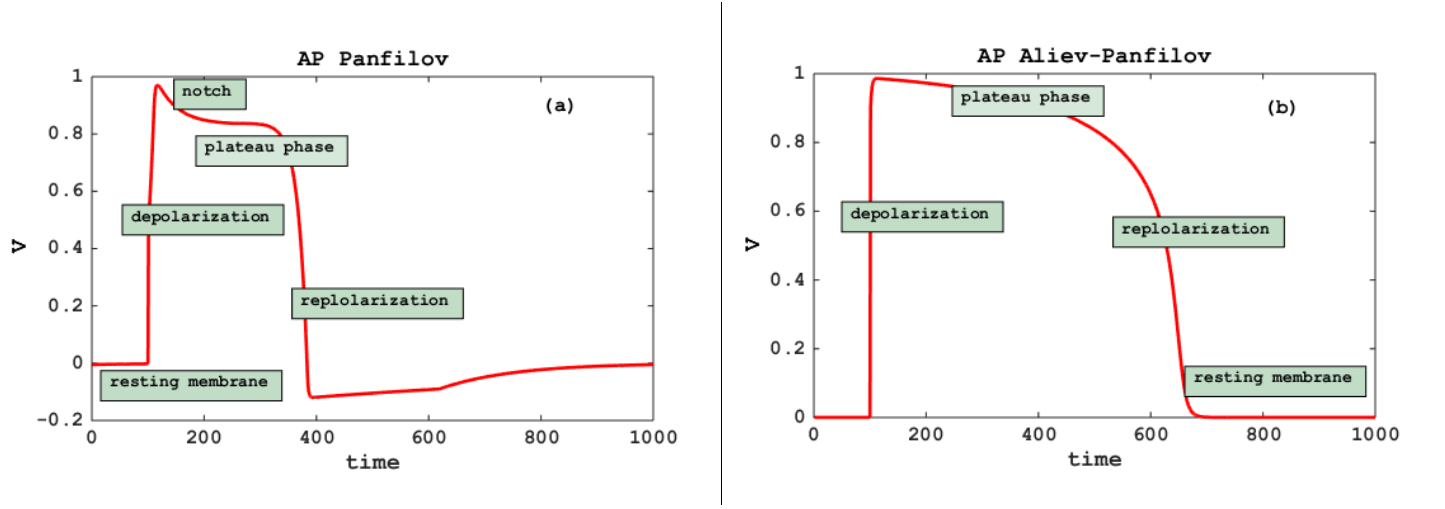}
\caption{Typical action potential shapes in (a) the Panfilov model and
(b) the Aliev-Panfilov model. The action potential duration (APD) for the
Panfilov model is $5.92$ time units (time step of $0.022$) and for the
Aliev-Panfilov model is $42.12$ time units (time step of $ 0.06$).}
\label{ap}
\end{center}
\end{figure*}
%%%%%%%%%%%%%%%%%%%%%%%%%%%%%%%%%%%%%%%%%%%%%%%%% 
\begin{figure}
\begin{center}
\includegraphics[width=1\linewidth]{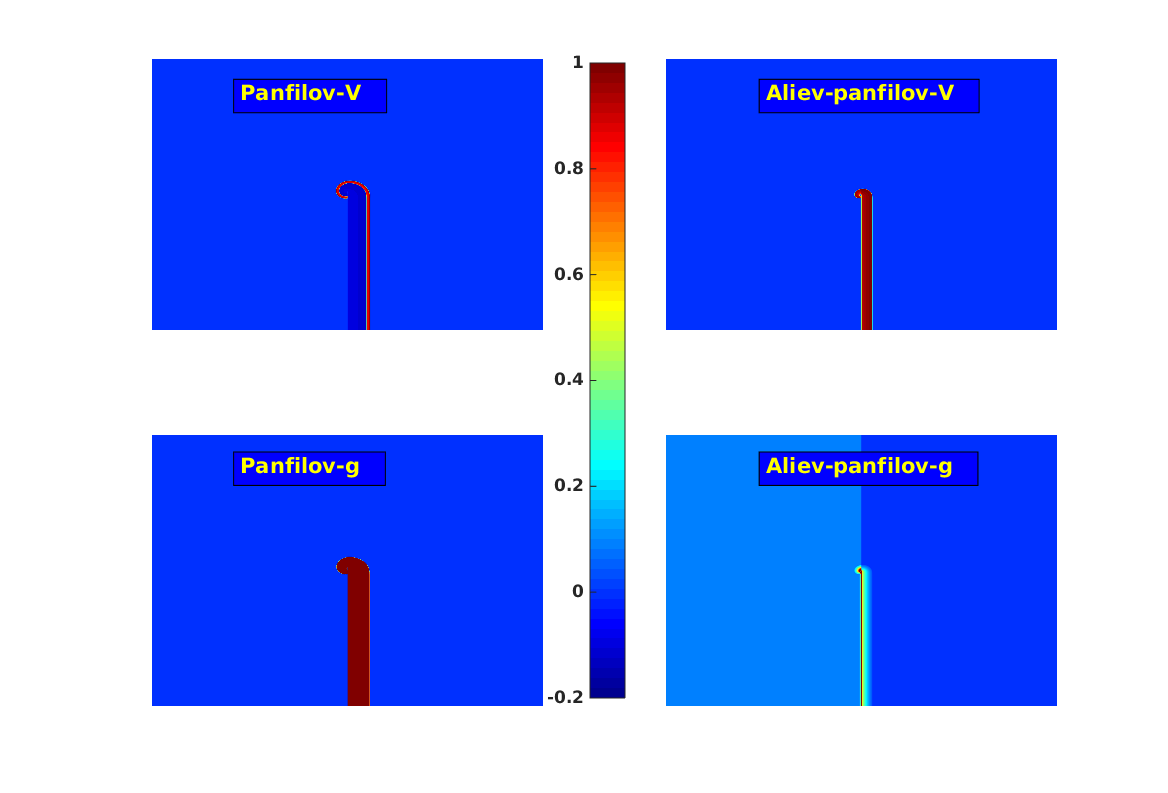}
\caption{Pseudocolor plots of the transmembrane potential $V$ (top panels) and
the recovery variable $g$ (bottom panels) showing the spiral waves that we use
as initial conditions for our simulations of the 2D Panfilov (left panels) and
the Aliev-Panfilov (right panels) models. }
\label{ini_2d}
\end{center}
\end{figure}
%%%%%%%%%%%%%%%%%%%%%%%%%%%%%%%%%%%%%%%%%%%%%%%%%%%%%%%%%%%
\begin{figure*}
\begin{center}
\includegraphics[width=1\linewidth]{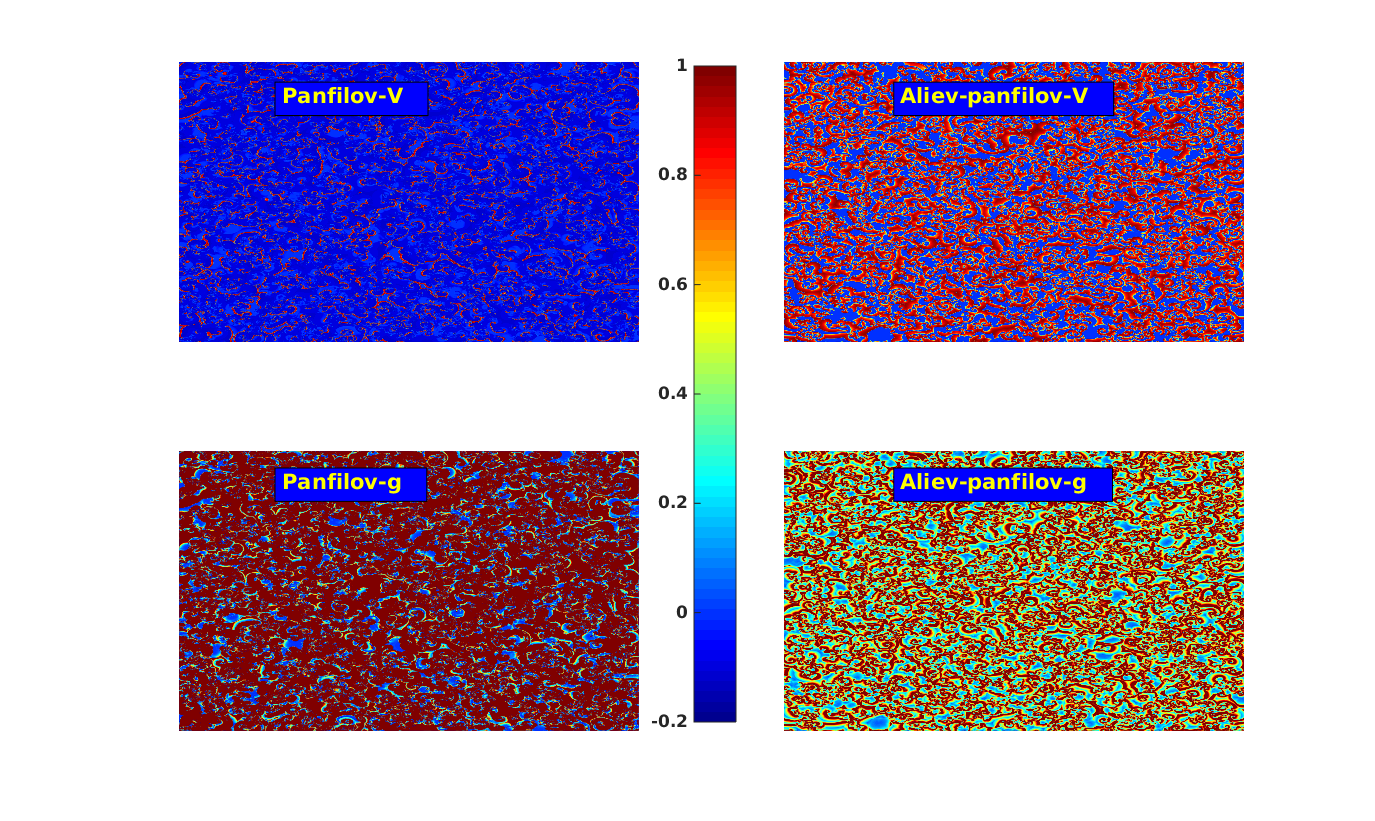}
\caption{Pseudocolor plots of the transmembrane potential $V$ (top panels) and
the recovery variable $g$ (bottom panels) showing typical turbulent states in
our simulations of the 2D Panfilov (left panels) and the Aliev-Panfilov (right
panels) models at a representative time, $7.5$ (dimensionless units), in the
turbulent, statistically steady state, which is, to a good approximation,
homogeneous and isotropic far away from boundaries. The spiral-wave initial
conditions of Fig.~\ref{ini_2d} have evolved into small, broken spirals that
interact with each other. For the complete spatio temporal
evolution of the spirals, see the videos S1,S2,S3 and S4 in the Supplementary
Material at~\cite{SuppMat}.}
\label{final_2d}
\end{center}
\end{figure*}
%%%%%%%%%%%%%%%%%%%%%%%%%%%%%%%%%%%%%%%%%%%%%%%%%%%%%%%%%%%
\begin{figure*}
\begin{center}
\includegraphics[width=1\linewidth]{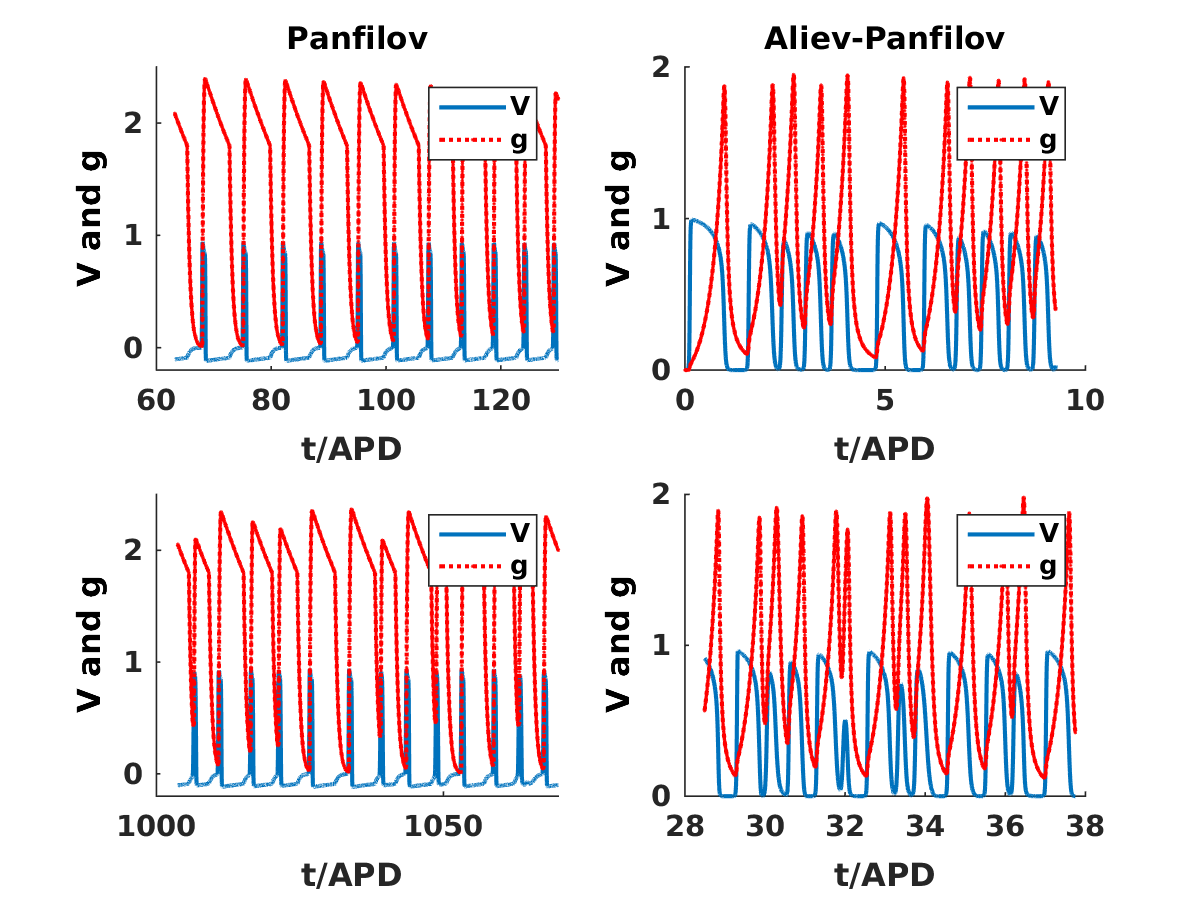}
\caption{Plots of the local time series of $V$ (blue lines) and $g$ (red,
dashed lines), obtained from the representative point $(x=2000, y=2000)$ (in
dimensionless units), from our 2D domains for the Panfilov (left panels) and
the Aliev-Panfilov (right panels) models; time is normalized by the Action
Potential Duration (APD), which is $5.92$ time units for the Panfilov model and
$42.12$ time units for the Aliev-Panfilov model (see Fig.~\ref{ap}). The domain
size is $4096 \times 4096$. The upper panels show the initial transients in the
time series; the lower panels show a section of these time series in the
statistically steady, spiral-wave-turbulence states.} 
\label{2dts}
\end{center}
\end{figure*}
%%%%%%%%%%%%%%%%%%%%%%%%%%%%%%%%%%%%%%%%%%%%%%%%%%%%%%%%%%%%%%%%%%%%%
\begin{figure}
\begin{center}
\includegraphics[width=1\linewidth]{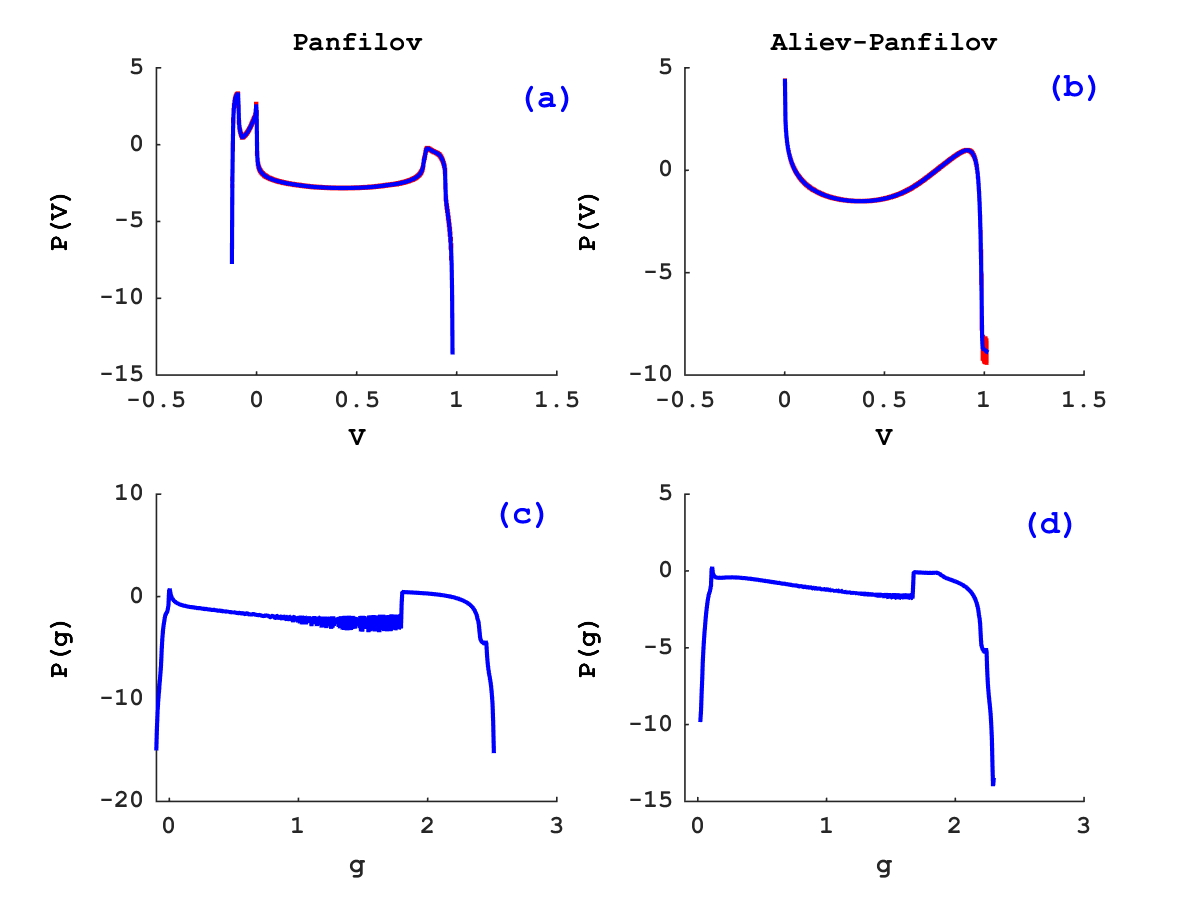}
\caption{Plots of PDFs of $V$ for (a) Panfilov and (b) Aliev-Panfilov models
and of $g$ for (c) Panfilov and (d) Aliev-Panfilov models in 2D. Here and
henceforth, in plots of PDFs, blue curves show the PDF and red symbols the error
bars. Error bars are often small and barely visible on the scales of these 
figures.}
\label{2dpdf_vg}
\end{center}
\end{figure}
%%%%%%%%%%%%%%%%%%%%%%%%%%%%%%%%%%%%%%%%%%%%%%%%%%%%%%%%

\section{Models, numerical methods and statistical properties}

We have used two simple, two-variable mathematical models for cardiac tissue,
namely, the Panfilov and the Aliev-Panfilov models. We describe these below. 

\subsection{Panfilov model}

The Panfilov model is a two-variable, Fitzgugh-Nagumo-type model, which is 
defined by the following partial differential equations (PDEs) for the 
transmembrane potential $V$ and the slow, recovery variable $g$:
\begin{eqnarray}
\frac{\partial V}{\partial t}&=&\nabla^2V-f(V)-g ;
\end{eqnarray}
\begin{eqnarray}
\frac{\partial g}{\partial t}&=\epsilon(V, g)(kV-g) . 
\end{eqnarray}\\
The initiation of the cardiac action potential is encoded in the following
functional form of $f(V)$: $f(V)=C_1V$, when $V<V_1$, $f(V)=-C_2V + a$, if
$V_1\leq V\leq V_2$ and $f(V)=C_3(V-1)$, for $V>V_2$. $\epsilon(V,
g)=\epsilon_1$ when $V<V_2$, $\epsilon(V, g)=\epsilon_2$ if $V>V_2$,
$\epsilon(V, g)=\epsilon_3$ for $V<V_1$ and $g<g_1$; here $V_1 = 0.0026, \, V_2
= 0.837, \, C_1 = 20, \, C_2 = 3, \, C_3 = 15, \, g_1 = 1.8, \, \epsilon_1 = 0.05, \, 
\epsilon_2 = 1, \, \epsilon_3 = 0.3, \, a = 0.06$ and $k = 3$. The
dynamics of the recovery variable is governed by $\epsilon(V, g)$: In
particular, $\epsilon_3^{-1}$ specifies the recovery time constants for small
$V$ and $g$ (its physiological equivalent is the relative refractory period);
$\epsilon_1^{-1}$ specifies the recovery time constant for relatively large
values of $g$ and intermediate values of $V$ (its physiological equivalent is
the absolute refractory period)~\cite{Panfilov-model}. We choose the values of
the parameters, which specify this model, as in
Ref.~\cite{Panfilov-model,shajahan}, so that we get experimentally reasonable
action potentials (APs). In this model, earlier
studies~\cite{Panfilov-model,shajahan} have obtained dimensioned time by
multiplying dimensionless time by $5$ ms so that the spiral-wave-rotation
period is $\simeq 120$ ms; and a space-scaling factor of $1$ mm yields a
spiral-arm wavelength of $\simeq 32.5$ mm; the dimensioned value of the
conductivity constant has been chosen to be $2$cm$^2$/s. In this paper, we
express all times in terms of the action-potential duration (APD). The action
potential (AP) for the Panfilov model is shown in Fig.~\ref{ap}(a). The action
potential duration (APD) for the Panfilov model is $5.92$ time units (time step
of $0.022$). 

\subsection{Aliev-Panfilov model}

Each site in the Aliev-Panfilov mathematical model for cardiac
tissue~\cite{Aliev-Panfilov-model} is described by the following coupled
equations for the transmembrane potential $V$ and the variable $g$, which
represents the conductance of the slow inward current: 
\begin{eqnarray}
\frac{dV}{dt} & = & -kV(V-a)(V-1)-Vg + I_{ex} ;
\label{ali1}
\end{eqnarray}
\begin{eqnarray}
\frac{dg}{dt} & = & [ \epsilon + \frac{\mu _1 g}{\mu _2 +V}] [ -g-kV(V-b-1)] ;
\label{ali2}
\end{eqnarray} 
$I_{ex}$ is the external current from all neighboring cells. The first term in
Eq.~(\ref{ali1}) determines the fast processes; the function $ \epsilon +
\frac{\mu _1 g}{\mu _2 +V} $ in Eq.~(\ref{ali2}) determines the slow, recovery
phase of the action potential. For a cell at the location $i, j, k$ in the
simulation domain, the total external current $I_{ex}$ is
\begin{eqnarray}
I_{ex} & = & G[(V_{i+1, j, k} - V_{i, j, k}) + (V_{i-1, j, k} - V_{i, j, k}) \nonumber \\
& & + (V_{i, j+1, k} - V_{i, j, k}) + (V_{i, j-1, k} - V_{i, j, k}) \nonumber \\
& & + (V_{i, j, k+1} - V_{i, j, k}) + (V_{i, j, k-1} - V_{i, j, k}) ].
\end{eqnarray}
The parameter values we use are(~\cite{Aliev-Panfilov-model}); $ \mu_1=0.11, \,
\mu_2=0.3, \, k=8, \, \epsilon=0.01, \, a=0.1, \, b=0.1,$ and $G=2.772$.  To
solve the ordinary differential equations in the Aliev-Panfilov model we use
the forward-Euler-central-difference scheme. The action potential (AP) for the
Aliev-Panfilov model is shown in Fig.~\ref{ap}(b).  The action potential
duration (APD) for the Aliev-Panfilov model is $42.12$ time units (time step of
$ 0.06$).

For two-dimensional (2D) simulations we use a $4096 \times 4096$ square domain
and for three-dimensional (3D) simulations we use a $ 640\times640\times640$
cube with the conventional Neumann conditions on the boundaries of our
simulation domains. To carry out simulations in such large domains we use
Graphics Processing Units (GPUs). We program these GPUs by using CUDA (Compute
Unified Device Architecture)~\cite{cuda-man}. We use the NVIDIA GPU K20C, with
$4$ compute nodes, for which we have developed optimized codes for the
solutions of the equations for the Panfilov and the Aliev-Panfilov models;
these codes are $4-5$ times faster than their counterparts on a desktop
computer with an Intel core-i7-4770, $3.4$ GHz system. 

We have conducted a detailed statistical analysis of the spiral- and
scroll-wave turbulence states in 2D and 3D, respectively, for the Panfilov and
the Aliev-Panfilov models. In addition to pseudocolor or isosurface plots of
$V$ and $g$, which show the spatiotemporal evolution of these fields, we have
obtained probability distribution functions (PDFs) of $V$ and $g$ and of their
gradients; for these PDFs we average data over our simulation domain and over
time, after discarding data from initial times during which transients decay
and a statistically steady state is established. The size of the domain, the
space step and time step used in our DNSs ($\delta x$ and $\delta t$), the
total run-time of the program ($\tau_{tot}$), and the transient time during
which data are discarded ($\tau_{trans}$) are given in Table~\ref{table1}. 

%%%%%%%%%%%%%%%%%%%%%%%%%%%%%%%%%%%%%%%%%%%%%%%%%%%%%%%%%%%%%%%%%%%%%%%%%%%
\begin{figure}
\begin{center}
\includegraphics[width=1\linewidth]{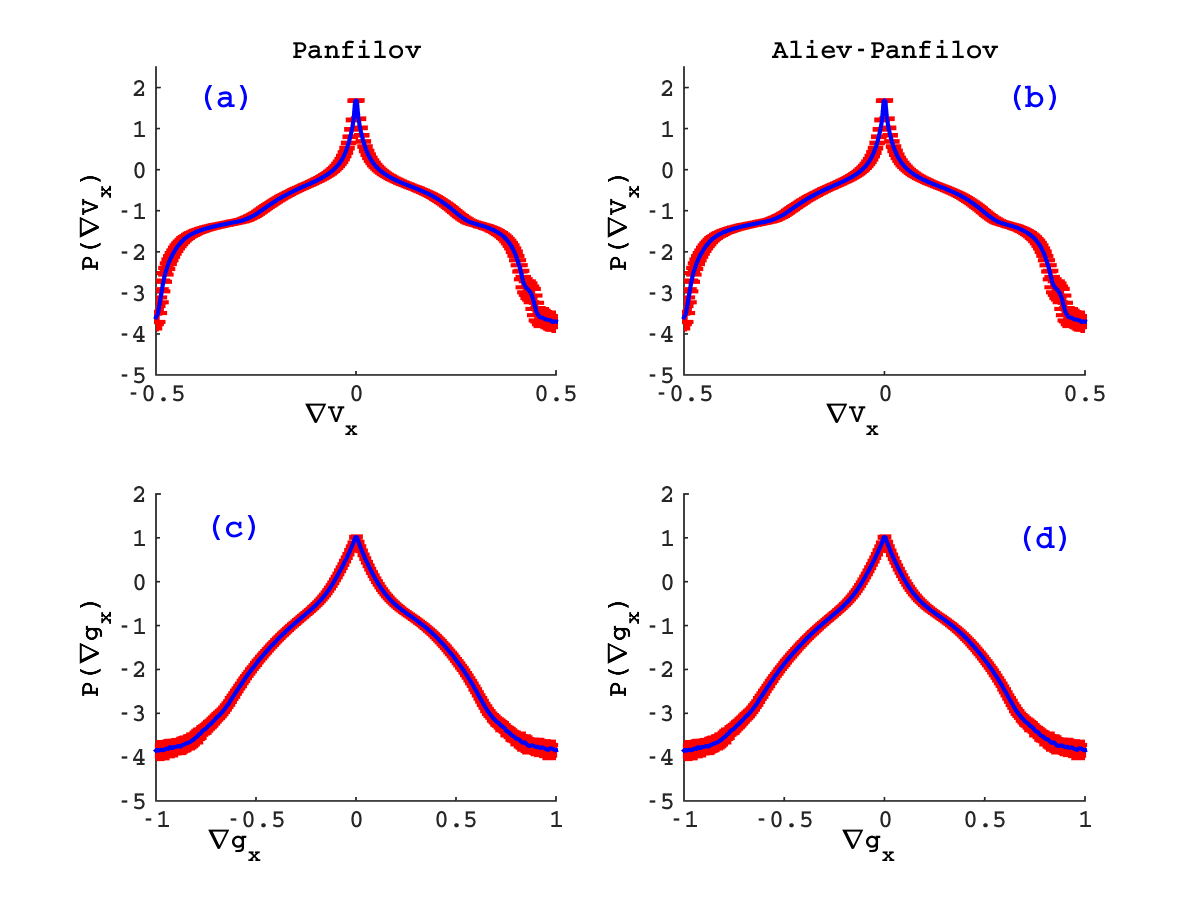}
\caption{Semilogarithmic plots (base 10) of PDFs of $(\nabla V)_x$ for (a)
Panfilov and (b) Aliev-Panfilov models and similar plots of PDFs of $(\nabla
g)_x$ for (c) Panfilov and (b) Aliev-Panfilov models in 2D.  These plots show
the components of these gradients are highly peaked at zero.  PDFs of $(\nabla
V)_y$ and $(\nabla g)_y$ show similar plots for the Panfilov and the
Aliev-Panfilov models.}
\label{2dpdf_gradvandg}
\end{center}
\end{figure}
%%%%%%%%%%%%%%%%%%%%%%%%%%%%%%%%%%%%%%%%%%%%%%%%%%%%%%%%%%%%%%%%%%%%%%%%%%%%%
\begin{table*}[]

\caption{The table shows, for each of our DNSs, the size of our simulation
domains, the space and time steps $\delta x$ and $\delta t$, respectively, the
total run-time of the program ($\tau_{tot}$) and the transient time during
which data are discarded ($\tau_{trans}$); numbers are in dimensionless space
and time units (see text).}
\begin{tabular}{p{4cm}p{4cm}p{1.7cm}p{1.7cm}p{1.7cm}p{1.7cm}}
	&	&	&	&	&	\\
\hline\hline
DNS &Domain size & $\delta x$ &$\delta t$& $\tau_{tot}$ & $\tau_{trans}$ \\
\hline \hline
2D-Panfilov & $4096\times4096$ & 0.5 & 0.022 & 13.2 & 7 \\
2D-Aliev-Panfilov & $4096\times4096$ & 0.5 & 0.06 & 12 & 6 \\
3D-Panfilov & $640\times640\times640$ & 0.5& 0.022 & 4.62 & 2.2 \\ %100 to 210 steps
3D-Aliev-Panfilov & $640\times640\times640$ & 0.5& 0.06 & 6 & 3 \\
\end{tabular}
\label{table1}
\end{table*}
%%%%%%%%%%%%%%%%%%%%%%%%%%%%%%%%%%%%%%%%%%%%%%%%%%%%%%%%%%%%%%%%%%%%%%%%%%%%

For representative results for pseudocolor and isosurface plots for fluid
turbulence, in 2D and 3D and for PDFs of the velocity components, the vorticity
and velocity gradients, we refer the reader to Refs.~\cite{frisch, lesieur,
pramanareview, sreenivasanantonia, boffettaecke, kaneda, ganapatimhd}.  We note
briefly that PDFs of velocity components are very close to Gaussian ones,
isosurfaces of the modulus of the vorticity $\omega$ show tubes at large values
of $|\omega|$ in 3D fluid turbulence, whereas pseudocolor plots of $\omega$ in
2D fluid turbulence show, in the absence of friction, the formation of large
vortices and antivortices, which are associated with the inverse cascade of
energy, from small to large length scales, that we discuss below. 

We also calculate the spatial power spectra of $V$ and $g$, which are 
\begin{eqnarray} 
	E_V(k) &=& \sum_{k-1/2\leq k'\leq k+1/2}|\tilde{V}(\mathbf{k})'|^2 \nonumber \\
	E_g(k) &=& \sum_{k-1/2 \leq k'\leq k+1/2}|\tilde{g}(\mathbf{k}')|^2 , 
\end{eqnarray} 
where the tildes denote spatial Fourier transforms and $\mathbf{k}$ and
$\mathbf{k}'$ are wave vectors with magnitudes $k$ and $k'$, respectively. To
eliminate the effects of the boundaries of our simulation domains, we evaluate
these Fourier transforms in the central parts of these domains (in 3D we use a
region with $512^3$ grid points and in 2D a region with $2048^2$ grid points).

The fluid-turbulence analogs of these spectra are the energy spectra. We recall
that, in 3D fluid turbulence, the fluid-energy energy spectrum $E(k) \sim
k^{-5/3}$ in the inertial range $ L^{-1} \ll k \ll \eta^{-1}_d$, where $L$ is
the large length at which energy is pumped into the system and $\eta_d$ the
Kolmogorov scale at which viscous dissipation becomes significant~\cite{frisch,
lesieur, pramanareview, sreenivasanantonia}. This power-law form of the energy
spectrum is associated with a forward cascade of energy, \`{a} la Richardson,
from the energy-injection scale $L$ to the dissipation scale $\eta_d$. The
spectral exponent $-5/3$ is a consequence of the simple-scaling phenomenology
of Kolmogorov~\cite{K41}, which is often denoted by K41; modern experiments and
numerical simulations suggest that this exponent is modified slightly because
of multiscaling corrections that arise from intermittency~\cite{frisch,
pramanareview, sreenivasanantonia, boffettareview}. Two-dimensional fluid
turbulence is qualitatively different from its 3D counterpart~\cite{frisch,
lesieur, pramanareview, boffettaecke}, as emphasized by
Kraichnan~\cite{fjortoff, kraichnan, leith, batchelor, lesieur}, because the 2D
Navier-Stokes equations also conserve the vorticity (the curl of the fluid
velocity) in the inviscid, unforced limit. A major consequence of this is that
the energy spectrum of homogeneous, isotropic fluid turbulence shows an inverse
cascade of energy, from the energy-injection length scale to even larger length
scales and a forward cascade of enstrophy (the mean square vorticity) from the
energy-injection scale to small length scales. The inverse cascade leads to a
scaling region in the energy spectrum $E(k) \sim k^{-5/3}$, whereas the forward
cascade yields $E(k) \sim k^{-3}$ (in the absence of any friction on the 2D
fluid). 
In studies of fluid turbulence, the energy spectrum $E(k)$ is used to define the integral length scale
\begin{equation}
L_I = \frac {\int (E(k)/k) dk} {\int E(k)dk}
\end{equation}
 and the Taylor-microscale length
\begin{equation} 
L_{\lambda} = \left[ {\frac{\int E(k)dk}{\int k^2 E(k) dk}}\right] ^{1/2}.
\end{equation}
We calculate the analogs of these lengths for our cardiac-tissue models by using the spectra $E_v(k)$ and $E_g(k) $.

Studies of the statistical properties of fluid turbulence also use order-$p$
structure functions of velocity increments~\cite{frisch, pramanareview};
therefore, we calculate their counterparts for $V$ and $g$. We illustrate this
for structure functions of $V$. We follow Ref.~\cite{prasad2D,prasad2D1} and
calculate $ {V'}({\bf{r}}) = {V} - \langle {V} \rangle_t$, where $\langle
\rangle_t$ denotes an average over time; we define 
\begin{equation}
S_p(r) =\langle \{[V'({\bf{r_c + r}}, t) - V'({\bf{ r_c}}, t)]\cdot{\bf{r}}/r
\}^p \rangle_{{\bf{r_c}} }, 
\label{spdef}
\end{equation}
where ${\bf{r}}$ has magnitude $r$ and ${\bf{r_c}}$ is an origin and $\langle
\rangle_{\bf{r_c} }$ denotes an average over time and the origin. The
fluid-turbulence analogs of $S_p(r)$ scale as $r^{\zeta_p}$ for $\eta_d \ll r
\ll L$. The K41 result is $\zeta_p = \zeta_p^{K41} = p/3$; experimental and
numerical studies show that simple scaling is replaced by multiscaling, i.e.,
the multiscaling exponents $\zeta_p$ deviate significantly from
$\zeta_p^{K41}$, especially for $p > 3$.  The multiscaling exponents $\zeta_p$
can be determined from the slopes of linear regions in log-log plots of
$S_p(r)$ versus $r$. In any finite-resolution study, such as a DNS, this linear
region is limited, so an accurate determination of $\zeta_p$ is a challenge,
particularly for large values of $p$. The range over which such exponents can
be fit can be enlarged by using the extended-self-similarity (ESS)
procedure~\cite{benzi, frischess} in which slopes of linear regions in log-log
plots of $S_p(r)$ versus $S_3(r)$ yield the exponent ratios $\zeta_p/\zeta_3$.
(In the fluid-turbulence context, $\zeta_3 = 1 $, by virtue of the von
K\'arm\'an-Howarth relation~\cite{lesieur}, so a determination of
$\zeta_p/\zeta_3$ yields $\zeta_p$.) We use this ESS procedure to look for
signatures of multiscaling the order-$p$ structure functions of $V$. Numerical
studies of 2D fluid turbulence suggest the velocity analogs of Eq.~\ref{spdef}
do not exhibit multiscaling, but the vorticity always
do~\cite{prasad2D,prasad2D1, boffettareview}. 

%%%%%%%%%%%%%%%%%%%%%%%%%%%%%%%%%%%%%%%%%%%%%%%%%%%%%%%%%%%%%%%%%%%%%%%%%%
\begin{figure}
\begin{center}
\includegraphics[width=1\linewidth]{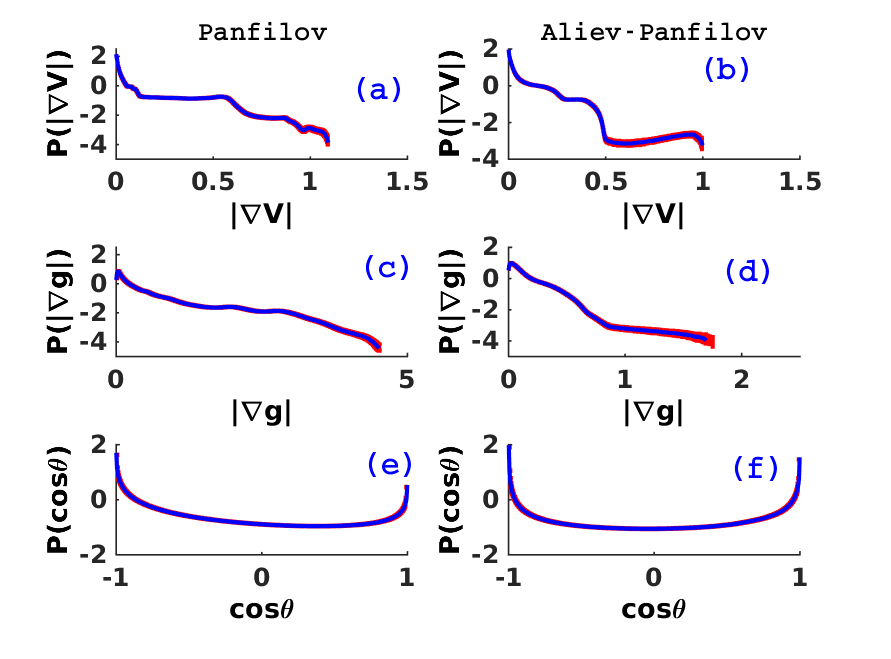}
\caption{Semilogarithmic (base 10) plots of the PDFs of (a) $|\nabla V| $, (c)
$| \nabla g|$, and (e) $\cos\theta = \frac{\nabla V . \nabla g}{| \nabla V |
|\nabla g |}$ for the 2D Panfilov model; (b), (d), and (f) show, respectively,
the corresponding PDFs for the 2D Aliev-Panfilov model. $|\nabla V|$ and
$|\nabla g|$ show dominant peaks at zero. Figures (e) and (f) show that $V$
and $g$ either tend to align or antialign with each other; the tendency
for being antialigned ($\cos\theta =-1$) is stronger than the tendency for
alignment ($\cos\theta = 1$).}
\label{2dpdf_modvg}
\end{center}
\end{figure}
%%%%%%%%%%%%%%%%%%%%%%%%%%%%%%%%%%%%%%%%%%%%%%%%%%%%%%%%%%%%%%%%%%%%%%%%%%
\begin{figure}
\begin{center}
\includegraphics[width=1\linewidth]{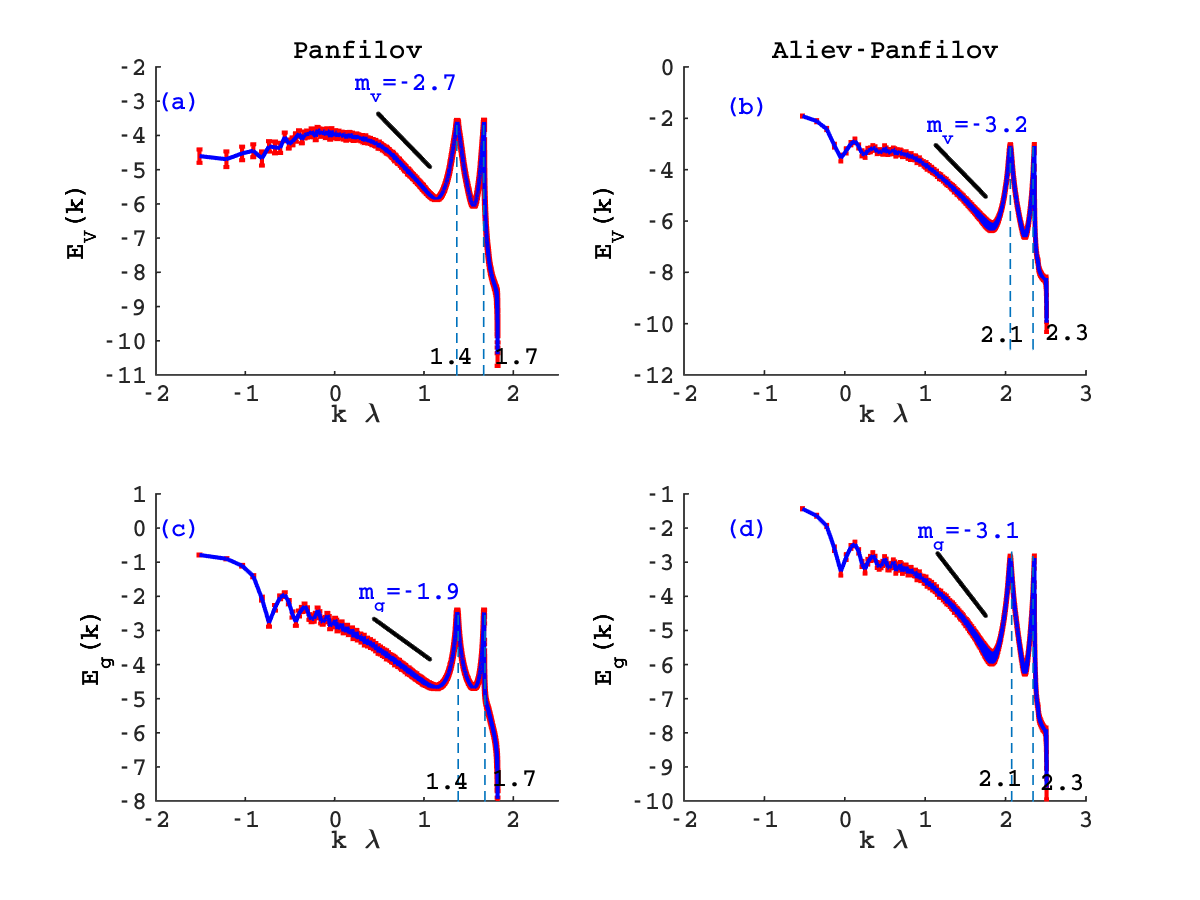}
\caption{Log-log (base 10) plots of the spectra $E_V(k)$ and $E_g(k)$ versus
$k\lambda $; here $\lambda=10$ dimensionless units for the Panfilov model and $
\lambda=48.5$ dimensionless units for the Aliev-Panfilov model, represents the
wavelength of a plane wave in the medium. The dashed, black line shows a
possible power-law regime. The overall shapes of these spectra are similar in
both models; in particular, we see two prominent peaks in these spectra,
located at $k\lambda = 1.4$ and $k\lambda = 1.7$ (Panfilov model) and at
$k\lambda = 2.1 $ and $ k\lambda = 2.4 $ (Aliev-Panfilov model).}
\label{2d_enerspec}
\end{center}
\end{figure}
%%%%%%%%%%%%%%%%%%%%%%%%%%%%%%%%%%%%%%%%%%%%%%%%%%%%%%%%%%%%%%%%%%%%%%%%%%%
\begin{figure}
\begin{center}
\includegraphics[width=1\linewidth]{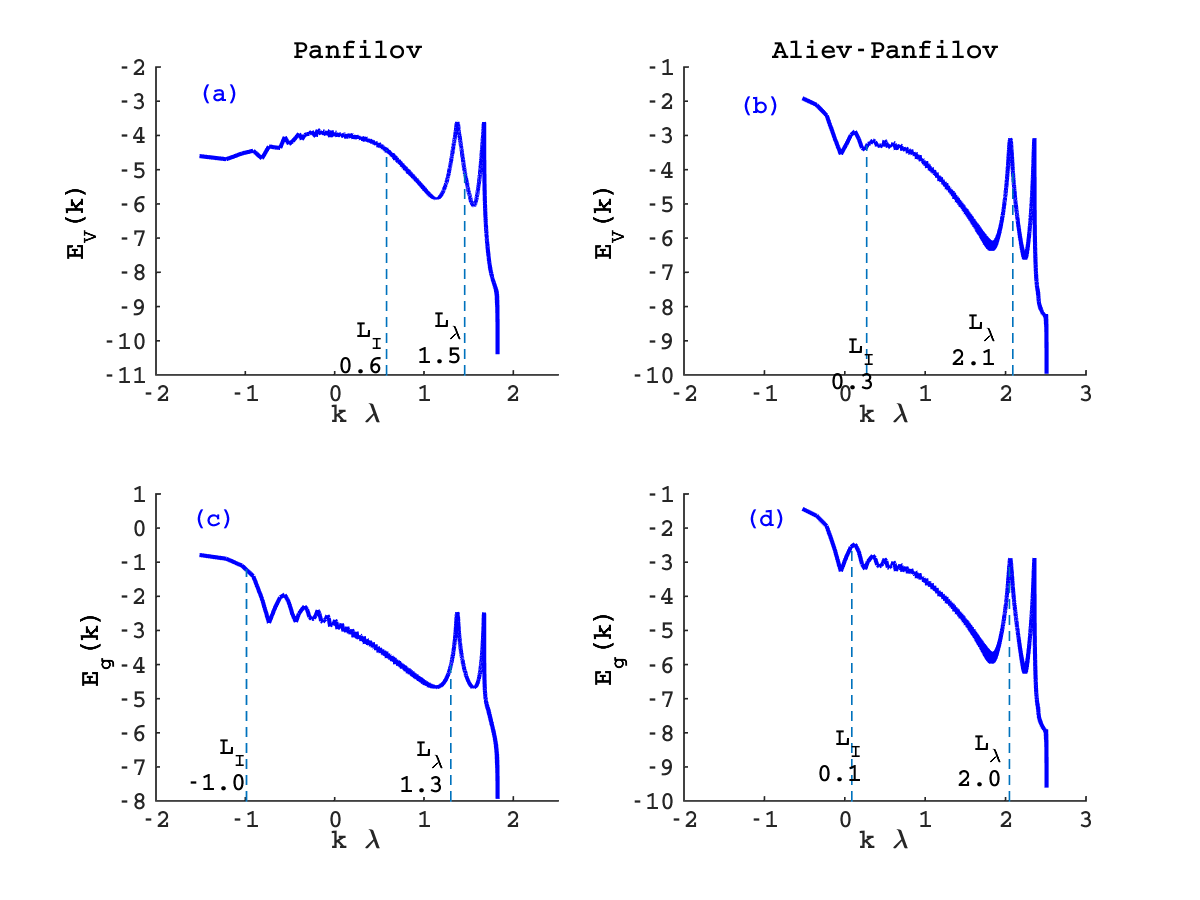}
\caption{Log-log (base 10) plots of the spectra $E_V(k)$ and $E_g(k)$ versus
$k\lambda $, with the positons of the Taylor micro length scale, $L_{\lambda}$ and the Integral length scale, $L_I$.}
\label{2dspectralengthscales}
\end{center}
\end{figure}

\begin{figure}
\begin{center}
\includegraphics[width=1\linewidth]{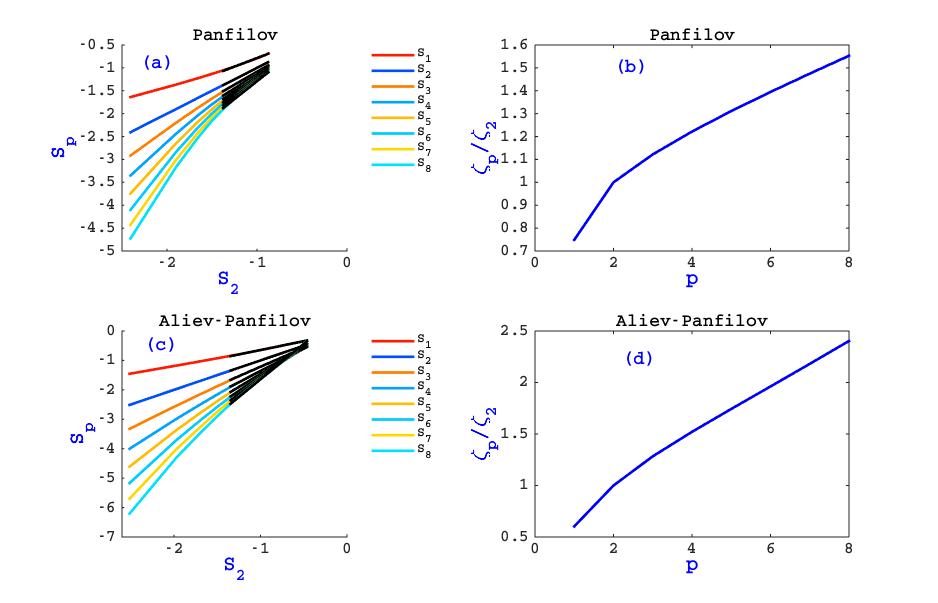}
\caption{Extended-self-similarity (ESS) plots of the structure functions $S_p$
of $V$ versus $S_2$ for 2D (a) Panfilov and (c) Aliev-Panfilov models, with $1
\leq p \leq 8$; the dashed (-) black lines represent possible power-low regimes.
In (b) and (d), we show plots versus $p$ of the exponent ratios
$\frac{\zeta_p}{\zeta_2}$ for the two models. The curvature of the plot of
$\frac{\zeta_p}{\zeta_2}$ versus $p$ indicates multiscaling.}
\label{2dstruct_ext}
\end{center}
\end{figure}
%%%%%%%%%%%%%%%%%%%%%%%%%%%%%%%%%%%%%%%%%%%%%%%%%%%%%%%%%%%%%%%%%%%%%%%%%%%%%%%
\begin{figure}
\begin{center}
\includegraphics[width=1\linewidth]{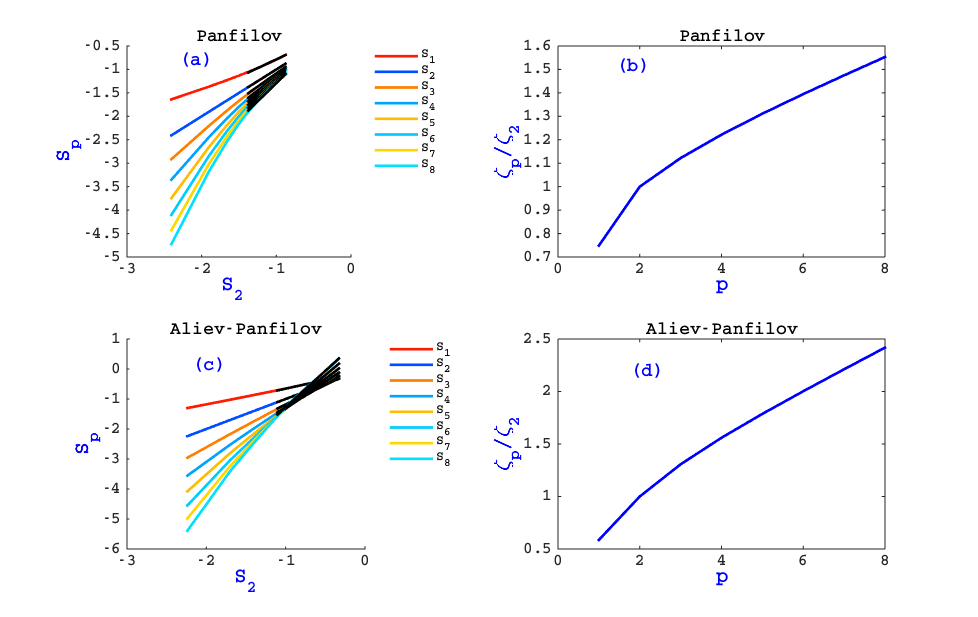}
\caption{Extended-self-similarity (ESS) plots of the structure functions $S_p$
of $g$ versus $S_2$ for 2D (a) Panfilov and (c) Aliev-Panfilov models, with $1
\leq p \leq 8$; the dashed (-) black lines represent possible power-low regimes.
In (b) and (d), we show plots versus $p$ of the exponent ratios
$\frac{\zeta_p}{\zeta_2}$ for the two models. The curvature of the plot of
$\frac{\zeta_p}{\zeta_2}$ versus $p$ indicates multiscaling.}
\label{2dstruct_extg}
\end{center}
\end{figure}
%%%%%%%%%%%%%%%%%%%%%%%%%%%%%%%%%%%%%%%%%%%%%%%%%%%%%
%%%%%%%%%%%%%%%%%%% 3D Figures %%%%%%%%%%%%%%%%%%%%%%%%
%%%%%%%%%%%%%%%%%%%%%%%%%%%%%%%%%%%%%%%%%%%%%%%%%%%%%%%%%%%%%%%%%
\begin{figure}
\begin{center}
\includegraphics[width=1\linewidth]{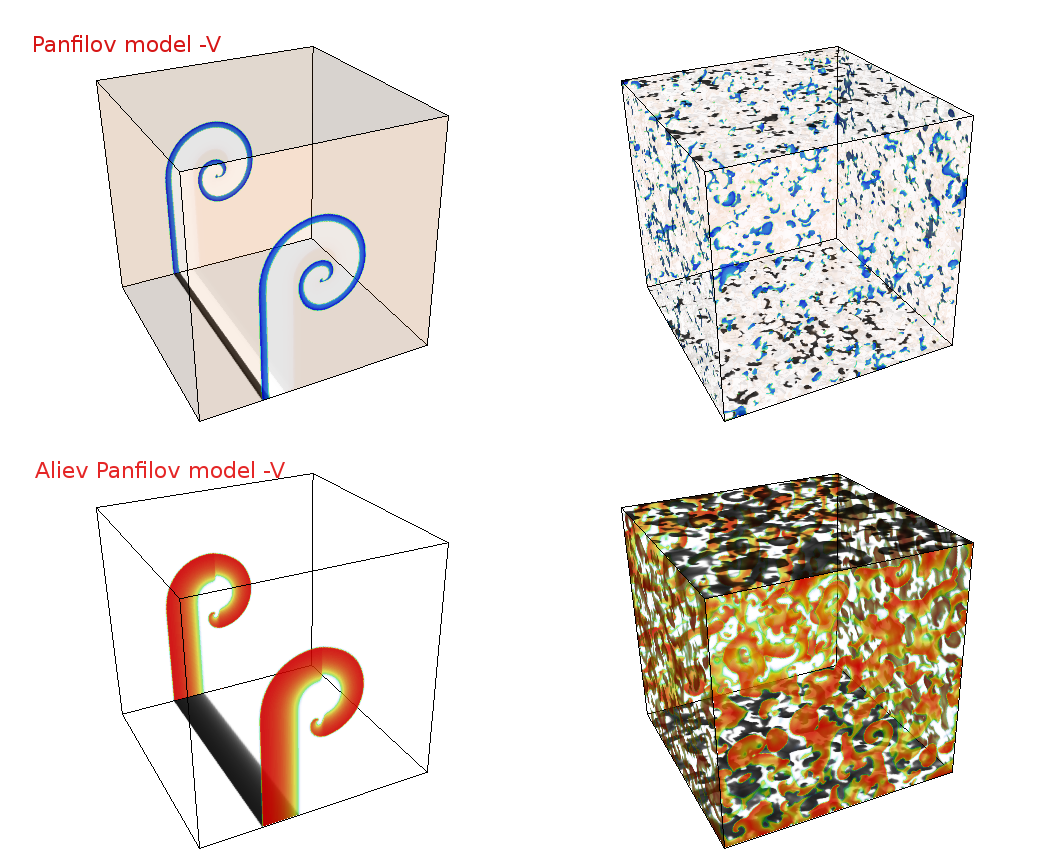}
\caption{Pseudocolor plots of the transmembrane potential $V$ showing the
scroll waves that we use as initial conditions for our simulations of the 3D
Panfilov (top, left panel) and the Aliev-Panfilov (bottom, left panel) models
and typical turbulent states in our simulations of the 3D Panfilov (top, right
panel) and Aliev-Panfilov (bottom, right panel) models, at a representative
time, $t=3.8$ (dimensionless units), in the turbulent, statistically steady 
state, which is, to a good approximation, homogeneous and isotropic far away 
from boundaries. For the complete spatio temporal evolution of the scrolls, see 
movies S5 and S6 in the Supplementary Material at~\cite{SuppMat}.}
\label{3dsnap_p}
\end{center}
\end{figure}
%%%%%%%%%%%%%%%%%%%%%%%%%%%%%%%%%%%%%%%%%%%%%%%%%%%%%%%%%%%
\begin{figure}
\begin{center}
\includegraphics[width=1\linewidth]{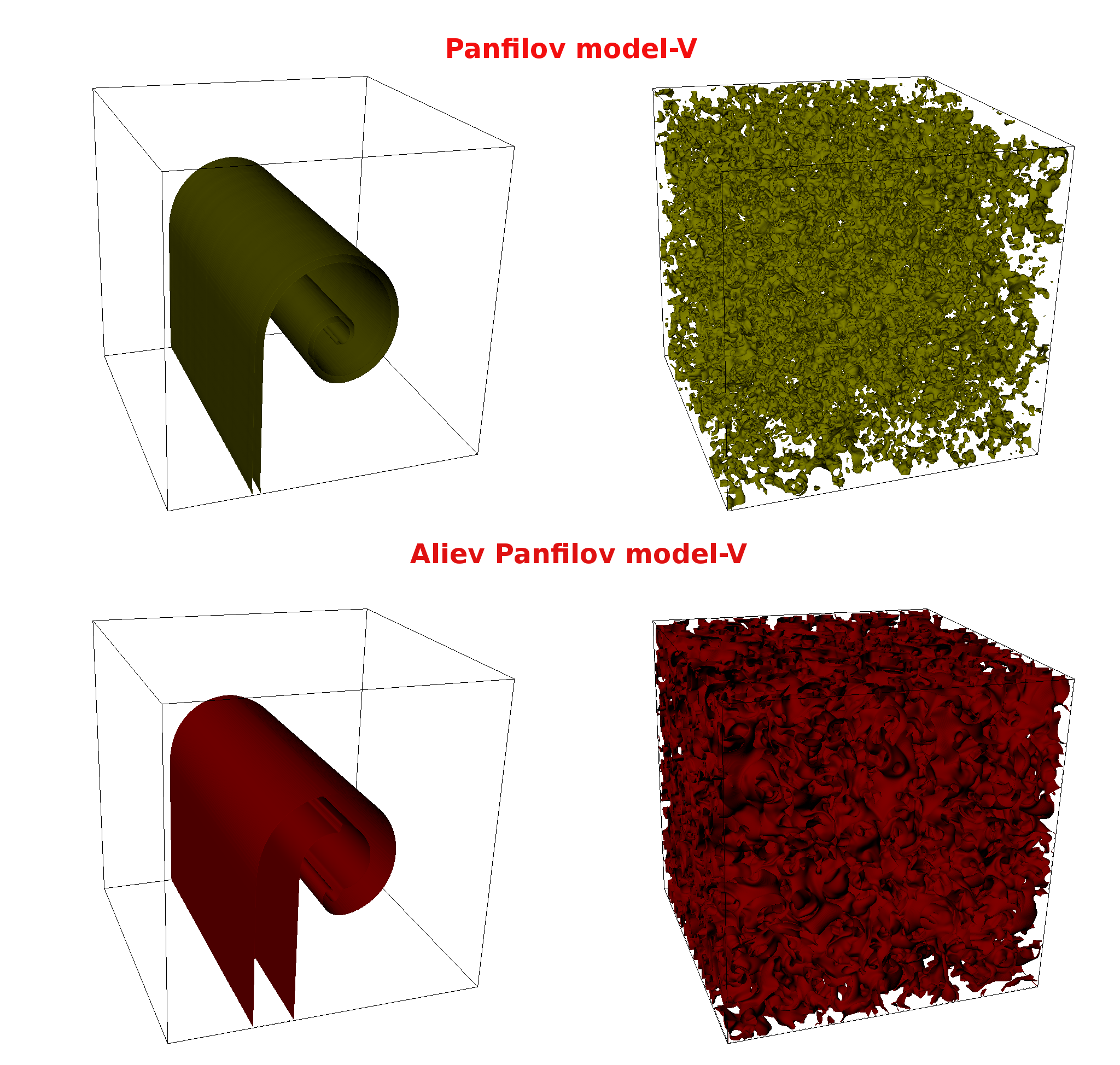}
\caption{Isosurface plots, for the 3D Panfilov (top panels) and Aliev-Panfilov
(bottom panels) models, of the transmembrane potential $V$ (isosurfaces with $V
= 0.8$), showing the scroll waves that we use as initial conditions for our
simulations of the 3D Panfilov (top, left panel) and the Aliev-Panfilov
(bottom, left panel) models and typical turbulent states in our simulations of
the 3D Panfilov (top, right panel) and Aliev-Panfilov (bottom, right panel)
models, at a representative time, $3.8$ (dimensionless units), in the turbulent,
statistically steady state, which is, to a good approximation, homogeneous and
isotropic far away from boundaries. For the complete spatio temporal evolution of the scrolls, 
 see movies S7 and S8 in the Supplementary Material at~\cite{SuppMat}.} 
\label{3dsnap_c}
\end{center}
\end{figure}
%%%%%%%%%%%%%%%%%%%%%%%%%%%%%%%%%%%%%%%%%%%%%%%%%%%%%%%%%%%

\begin{figure*}
\begin{center}
\includegraphics[width=1\linewidth]{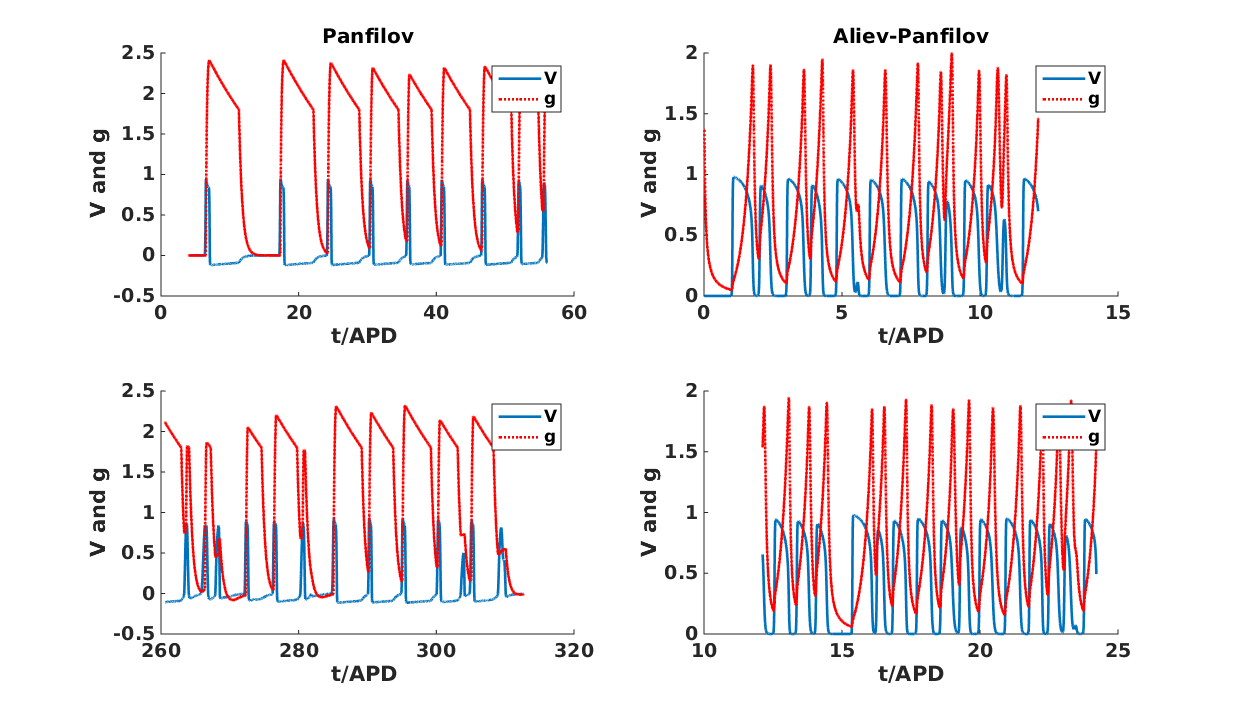}
\caption{Local time series of $V$ (blue lines) and $g$ (red dashed lines),
obtained from the representative point $(x=200, y=200, z=300)$ (in
dimensionless units) from our 3D domains for the Panfilov (left panels) and the
Aliev-Panfilov (right panels) models; time is normalized with the 
action-potential duration (APD); APD$=5.92$ time units for the Panfilov model;
APD$=42.12$ time units for the Aliev-Panfilov model. The domains size is $640
\times 640 \times 640$. The upper panels show the initial transients in the
time series; the lower panels show a section of these time series in the
statistically steady, scroll-wave-turbulence states. The time series of $V$
consists of a train of action potentials showing the states of depolarization,
plateau state, repolarization, and resting state of the excitable medium. The
action potential for the Aliev-Panfilov model does not have an overshoot
region, where the system repolarizses beyond its resting state.}\label{3dts}
\end{center}
\end{figure*}
\begin{figure}
\begin{center}
\includegraphics[width=1\linewidth]{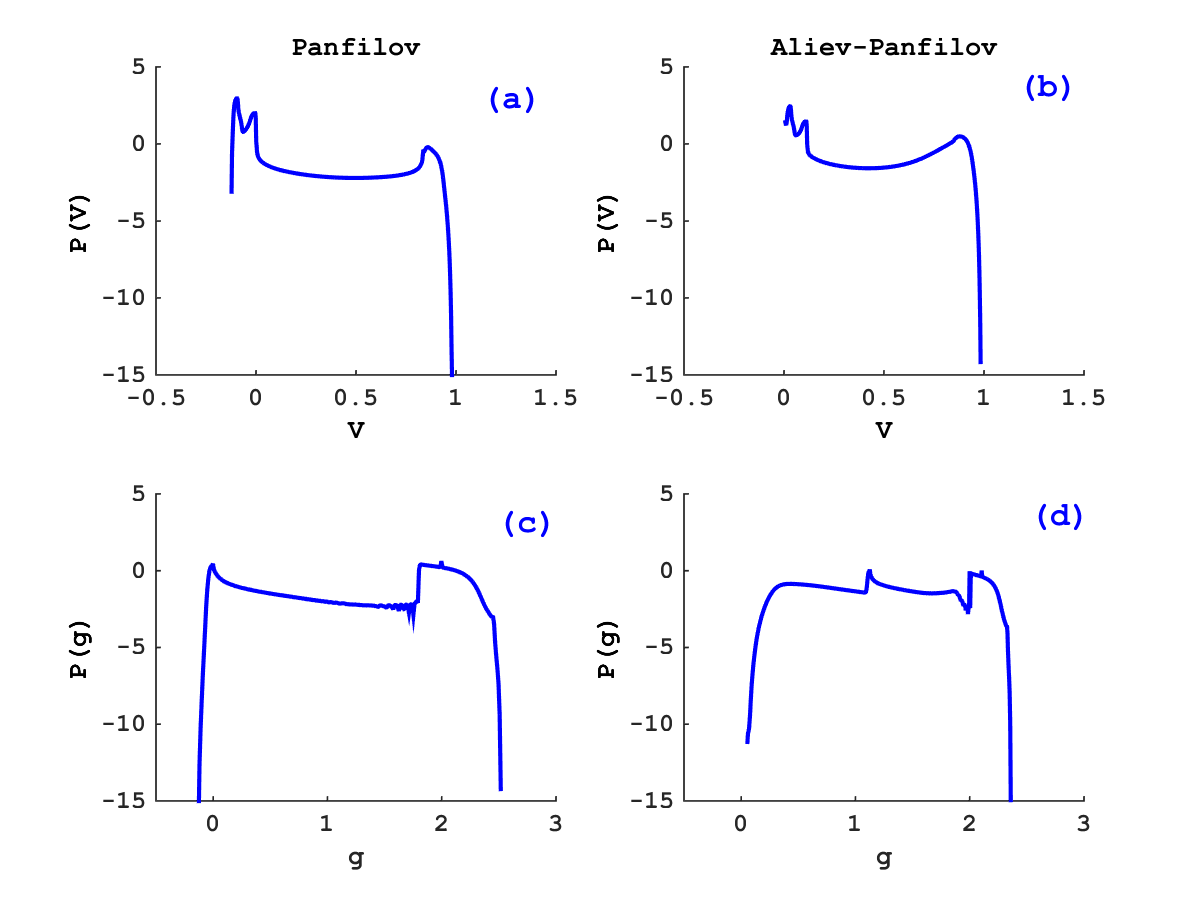}
\caption{Probability distribution functions (PDFs) of $V$ for (a) Panfilov 
and (b) Aliev-Panfilov models and of $g$ for (c) Panfilov and (d) 
Aliev-Panfilov models in 3D.}
\label{3dpdf_vg}
\end{center}
\end{figure}
%%%%%%%%%%%%%%%%%%%%%%%%
\begin{figure}
\begin{center}
\includegraphics[width=1\linewidth]{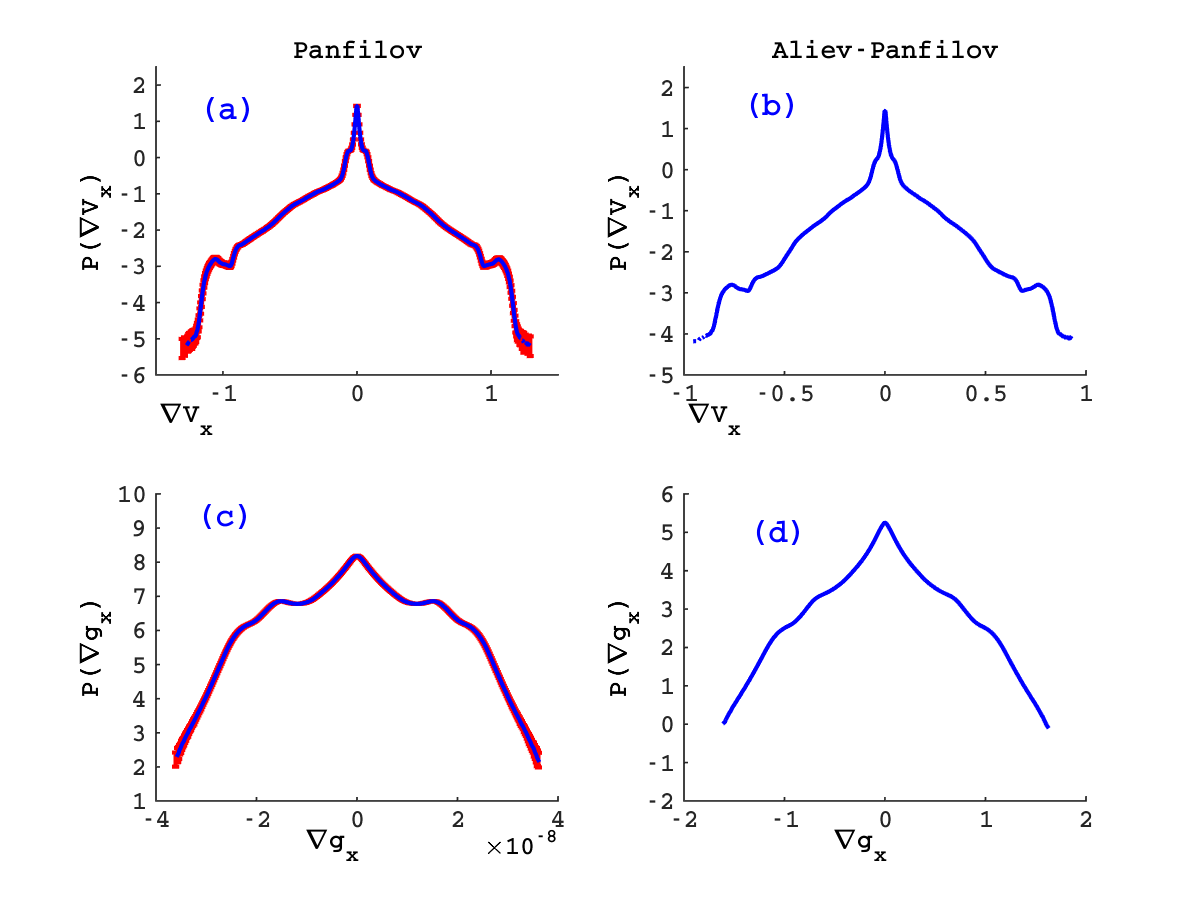}
\caption{Semilogarithmic (base-10) plots of PDFs of $(\nabla V)_x$ for (a)
Panfilov and (b) Aliev-Panfilov models and $(\nabla g)_x$ for (c) Panfilov and
(d) Aliev-Panfilov models in 3D. These plots show dominant peaks at zero. The
PDFs of $(\nabla V)_y$ and of $(\nabla V)_z$ are like those of
$(\nabla V)_x$; and PDFs of $(\nabla g)_y$ and of $(\nabla g)_z$ are akin
to those of $(\nabla g)_x$. }
\label{3dpdf_gradvandg}
\end{center}
\end{figure}
%%%%%%%%%%%%%%%%%%%%%%%%%%%%%%%%%%%%%%%%%%%%%%%%%%%%%%%%%%%%%%%%%%%%%%%%%%%%%%
\begin{figure}
\begin{center}
\includegraphics[width=1\linewidth]{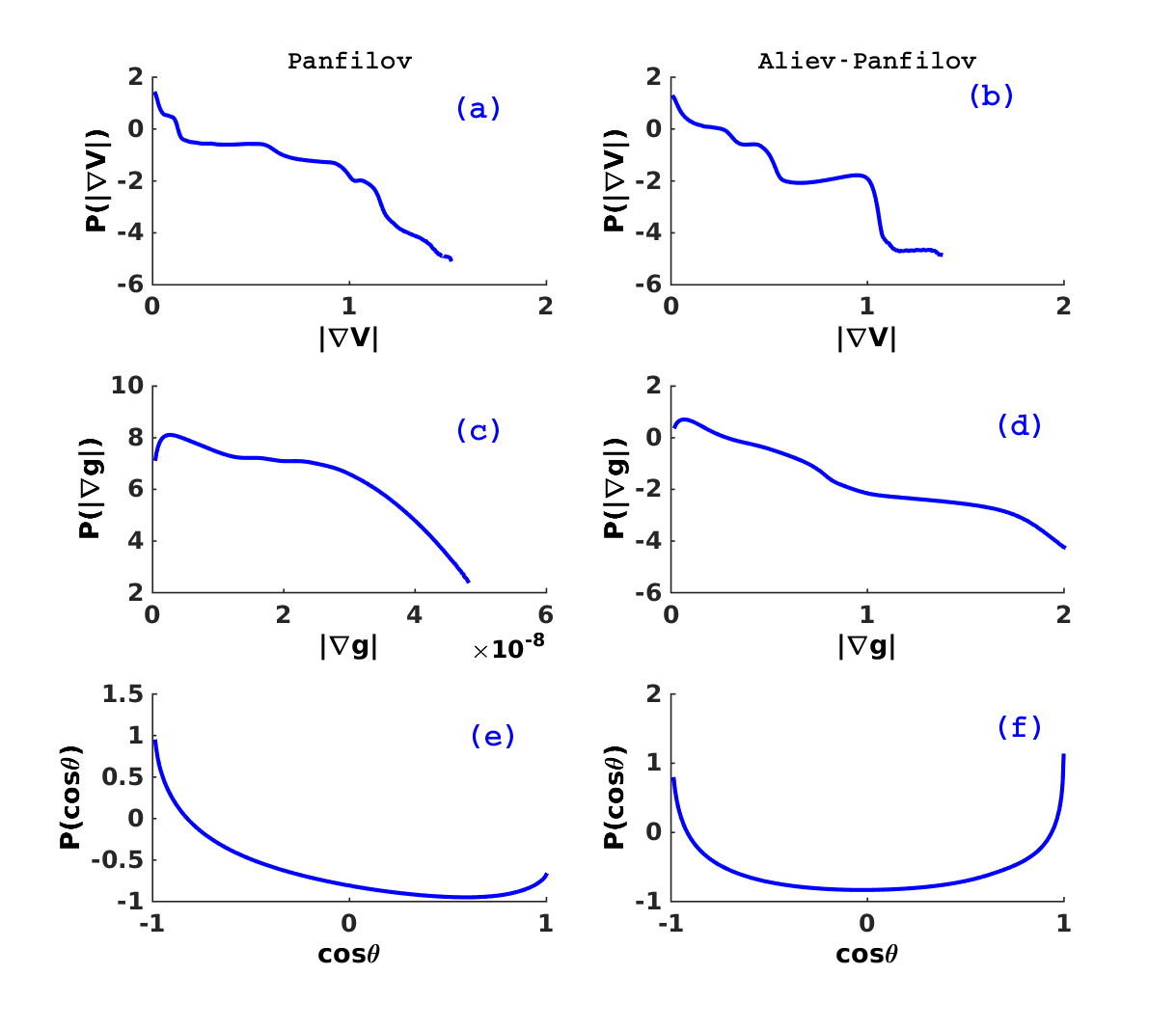}
\caption{Semilogarithmic (base 10) plots of the PDFs of (a) $|\nabla V|$, (c)
$| \nabla g|$ and (e)$\cos\theta = \frac{\nabla V . \nabla g}{|\nabla V |
|\nabla g |}$ for the 3D Panfilov model; (b), (d) and (f) show, respectively,
the corresponding PDFs for the 3D Aliev-Panfilov model; PDFs in (e) and (f)
show that $\nabla V$ and $\nabla g$ tend, on average, either to align or
anti-align with each other. The degree of alignment or anti-alignment is
different in these two models; for the Panfilov model, the tendency
for being antialigned($\cos\theta =-1$) is stronger than the tendency for
alignment ($\cos\theta = 1$).}\label{3dpdf_modvg}
\end{center}
\end{figure}
%%%%%%%%%%%%%%%%%%%%%%%%%%%%%%%%%%%%%%%%%%%%%%%%%%%%%%%%%%%%%%%%%%%%%%%%%%%%%%%%%
\begin{figure}
\begin{center}
\includegraphics[width=1\linewidth]{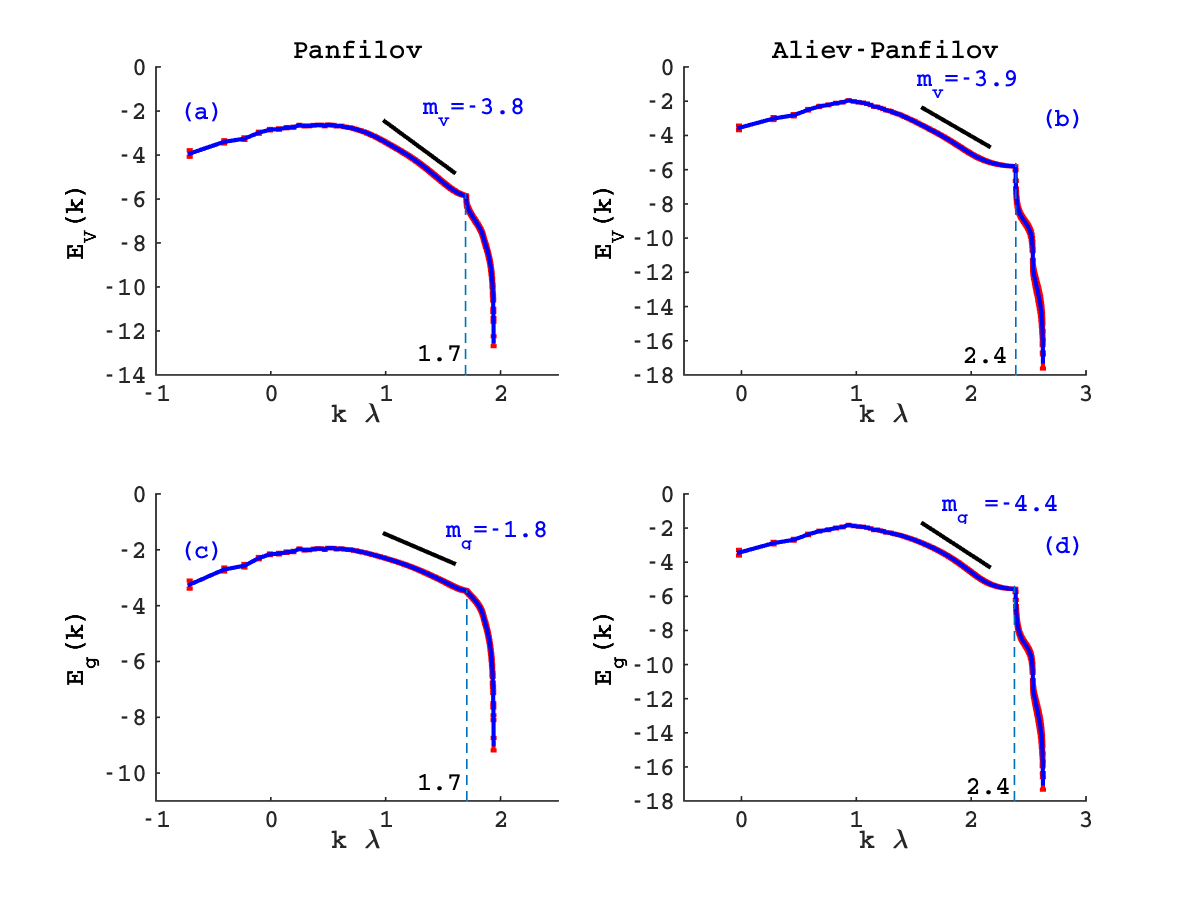}
\caption{Log-log (base 10) plots of the spectra $E_V(k)$ and $E_g(k)$ versus
$k\lambda $. Here $\lambda=10$, in dimensionless units for the Panfilov model, 
and $\lambda=48.5$, in dimensionless units for the Aliev-Panfilov model, 
represents the wavelength of a plane wave in the medium. The dashed, black 
lines show (possible) power-law regimes. The overall shapes of these spectra 
are similar in both models. We see one prominent peak in these spectra. 
This peak is located at $k\lambda = 1.7$ (Panfilov model) and $ k\lambda = 2.4
$ (Aliev-Panfilov model).}
\label{3d_enerspec}
\end{center}
\end{figure}
%%%%%%%%%%%%%%%%%%%%%%%%%%%%%%%%%%%%%%%%%%%%%%%%%%%%%%%%%%%%%%%%%%%%%%%%%%%%%%%
\begin{figure}
\begin{center}
\includegraphics[width=1\linewidth]{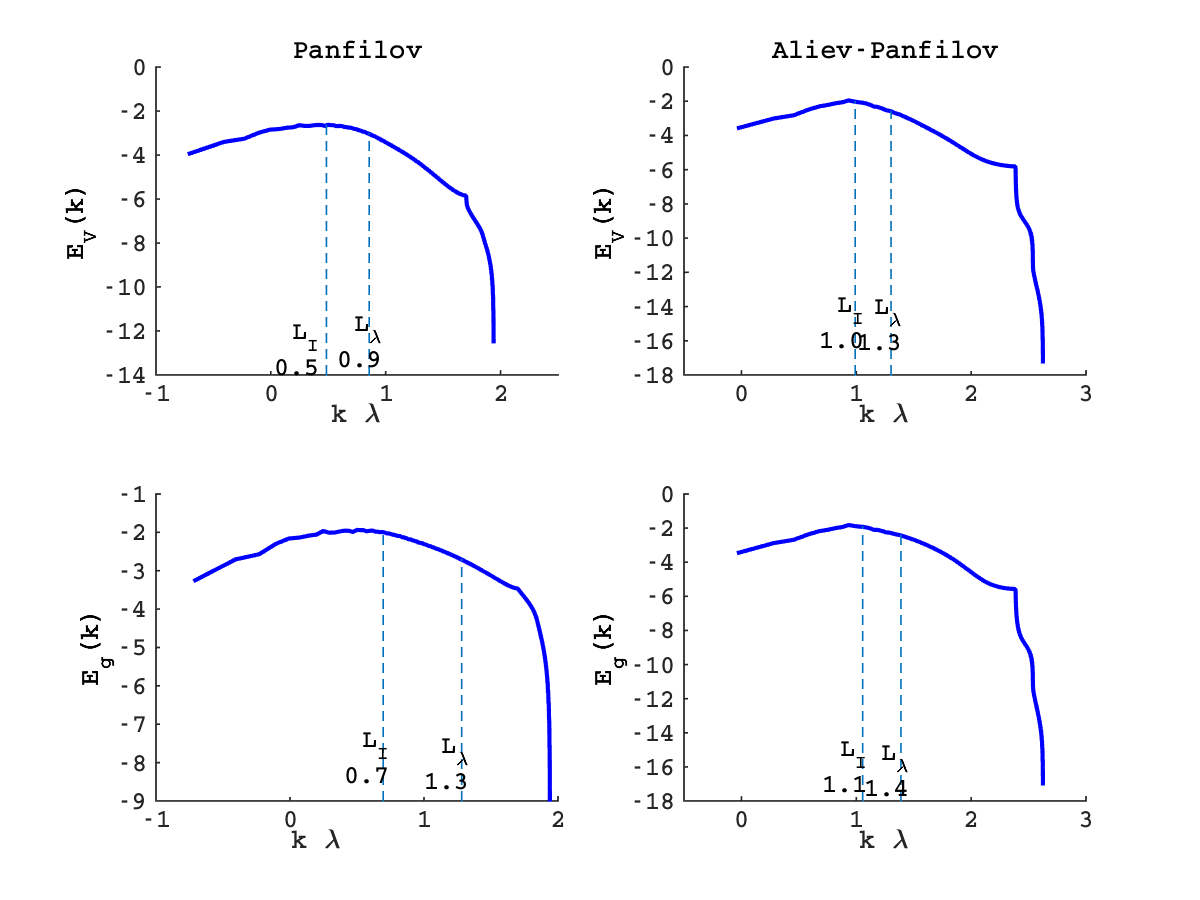}
\caption{Log-log (base 10) plots of the spectra $E_V(k)$ and $E_g(k)$ versus
$k\lambda $, with the positons of the Taylor micro length scale, $L_{\lambda}$ and the Integral length scale, $L_I$.}
\label{3dspectralengthscales}
\end{center}
\end{figure}
\begin{figure}
\begin{center}
\includegraphics[width=1\linewidth]{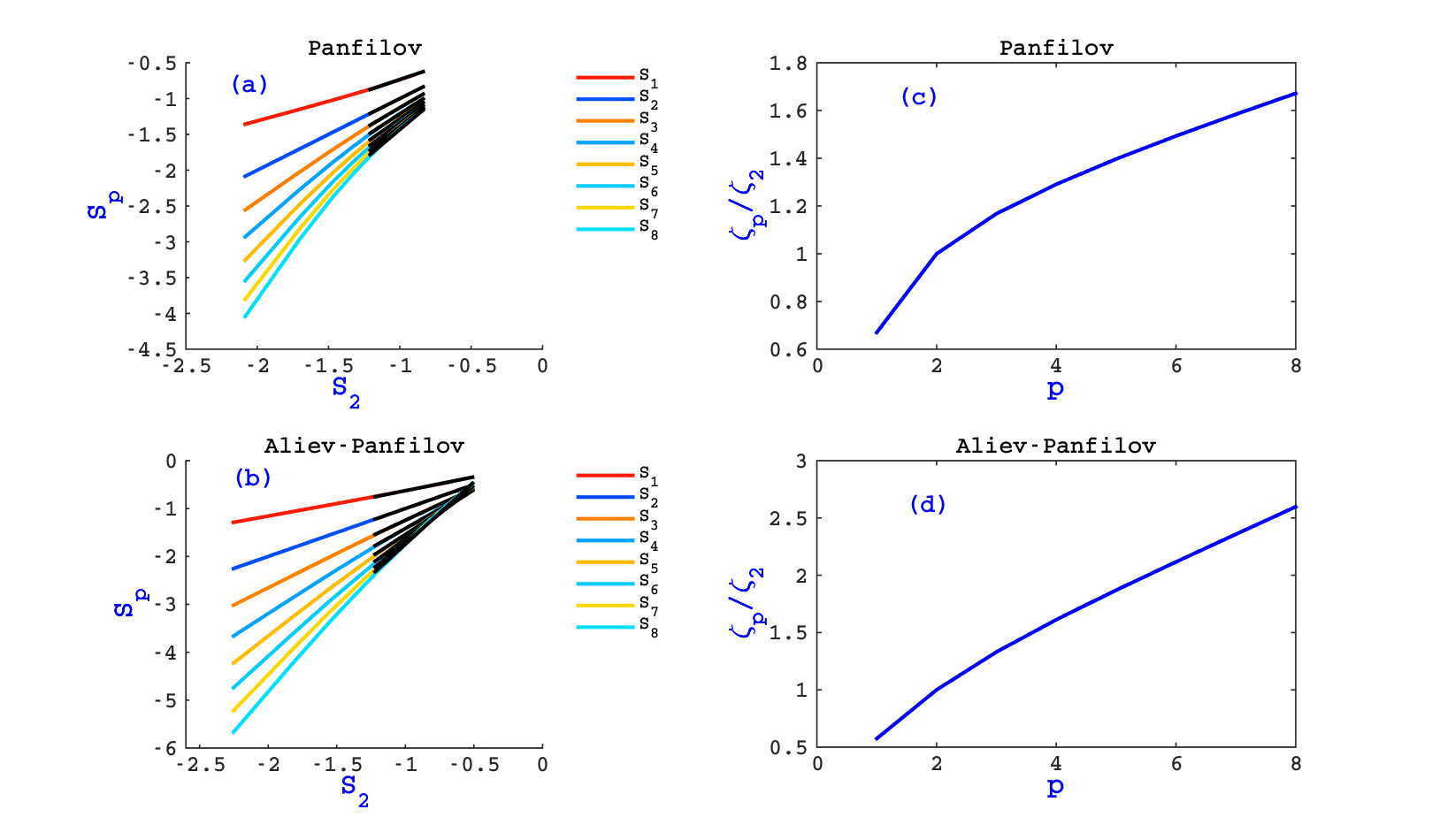}
\caption{Extended-self-similarity (ESS) plots of the structure functions $S_p$
of $V$ versus $S_2$ for 3D Panfilov (top, left panel) and
Aliev-Panfilov (bottom, left panel) models, with $1 \leq p \leq 8$; the
dashed-black lines indicate (possible) power-low regimes. In (b) and (d), we 
show plots of the exponent ratios $ \zeta_p/\zeta_2$ versus $p$ for the two 
models. The curvature of $\zeta_p/\zeta_2 $ versus $p$ indicates multiscaling.}
\label{3dstruct_ext}
\end{center}
\end{figure}
\begin{figure}
\begin{center}
\includegraphics[width=1\linewidth]{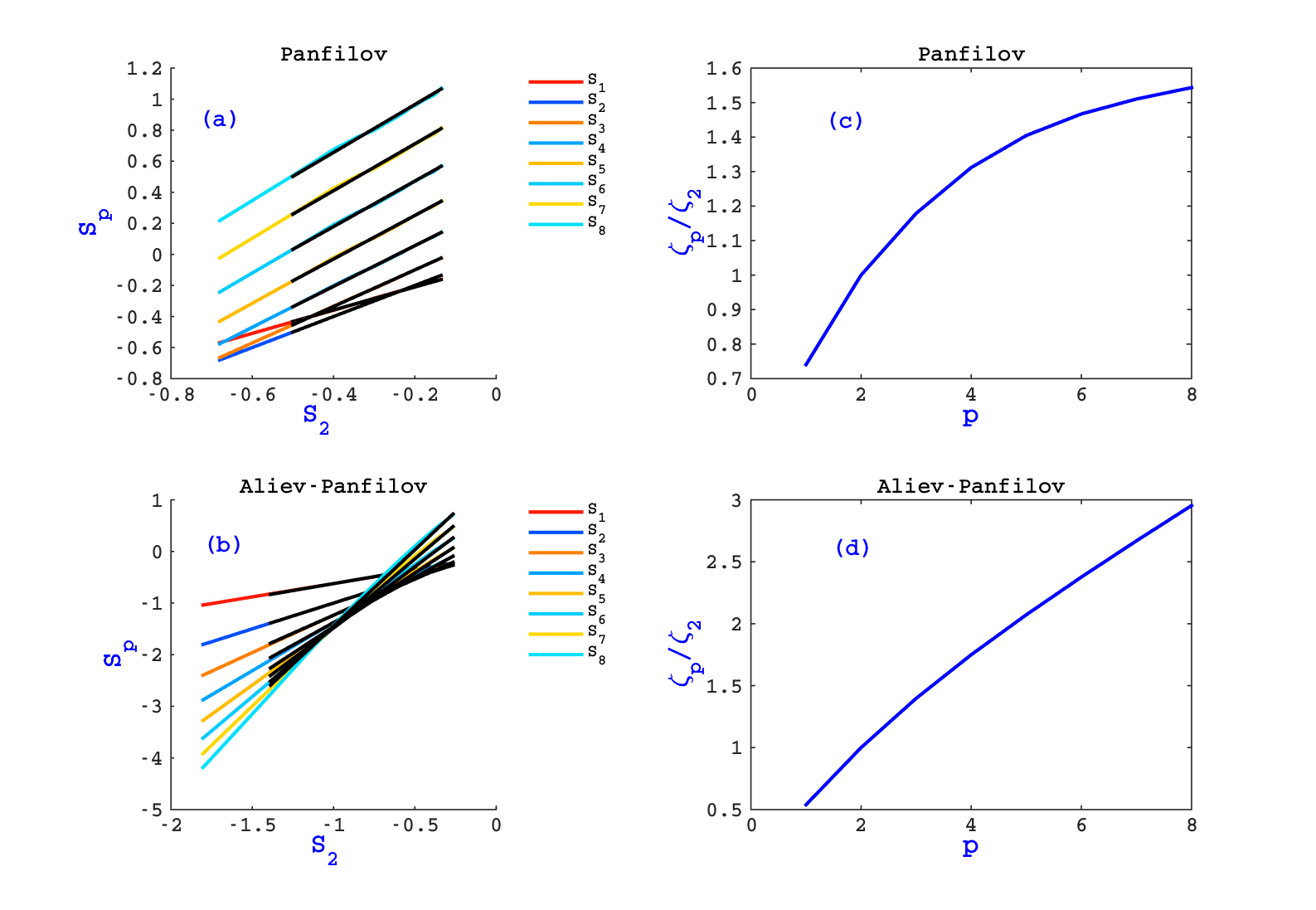}
\caption{Extended-self-similarity (ESS) plots of the structure functions $S_p$
of $g$ versus $S_2$ for 3D Panfilov (top, left panel) and
Aliev-Panfilov (bottom, left panel) models, with $1 \leq p \leq 8$; the
dashed-black lines indicate (possible) power-low regime. In (b) and (d), we show
plots of the exponent ratios $ \zeta_p/\zeta_2$ versus $p$ for the two models.
The curvature of $\zeta_p/\zeta_2 $ versus $p$ indicates multiscaling. }
\label{3dstruct_extg}
\end{center}
\end{figure}
%%%%%% End 3D Figures%%%%%%%%%%%%%%%%%%%%%%%
%%%%%%%%%%%%%%%%%%%%%%%%%%%%%%%%%%%%%%%%%%%%%%%%%%%%%%%%%%%%%%%%%%%%%%%%%%%%%%%%
\section{Results}

In this Section, we first present results from our simulations for spiral-wave
turbulence for the 2D Panfilov and the Aliev-Panfilov models. We then give
results from our studies of scroll-wave turbulence for these two models in 3D. 

\subsection{Results for two-dimensional spiral-wave turbulence}

Our DNSs in 2D use a domain of size $4096 \times 4096 $. We begin with half the
medium excited and the other half in the excitable state; and then we apply a
stimulus at the middle of the domain to form a spiral wave, which we use as an
initial condition. Figure~\ref{ini_2d} shows pseudocolor plots of the
transmembrane potential $V$ (top panels) and the recovery variable $g$ (bottom
panels) with the spiral waves that we use as initial conditions for our
simulations of the 2D Panfilov (left panels) and the Aliev-Panfilov (right
panels) models. Given our set of parameter values, these spiral waves break up
and the medium is filled with broken spiral waves. Figure~\ref{final_2d} shows
pseudocolor plots of the transmembrane potential $V$ (top panels) and the
recovery variable $g$ (bottom panels) that depict typical turbulent states in
our simulations of the 2D Panfilov (left panels) and the Aliev-Panfilov (right
panels) models at a representative time, $7.5$ (dimensionless units), in the
turbulent, statistically steady state, which is, to a good approximation,
homogeneous and isotropic far away from boundaries. Note that the spiral-wave
initial conditions of Fig.~\ref{ini_2d} evolve into small, broken spirals
that interact with each other. This is the analog of the Richardson cascade in
fluid turbulence. Instead of large eddies breaking down into ever smaller and
lighter eddies, here a large spiral wave breaks down into smaller and smaller
spiral waves.  These pseudocolor plots can be contrasted with their analogs for
the vorticity in 2D fluid turbulence~\cite{prasad2D}, where the typical sizes
of vortical regions are governed by (a) the energy-injection length scale and
(b) whether or not there is air-drag-induced friction on the 2D fluid. 

After the system reaches a statistically steady state, we record the values of
$V$ and $g$; and we then calculate their statistical properties, which we
describe below. 

\subsection{Statistical properties of spiral-wave turbulence} 

Figure~\ref{ap} shows typical action potentials for the Panfilov (top panel)
and the Aliev-Panfilov (bottom panel) models, illustrating the depolarization,
plateau, repolarization, and resting states of the excitable medium. We present
our results in Figs.~\ref{2dts}-\ref{2dstruct_extg}. Figure~\ref{2dts} shows
local time series of $V$ (blue lines) and $g$ (red dashed lines), obtained from
the representative point $(x=2000, y=2000)$ (in dimensionless units) from our
2D domains for the Panfilov (left panels) and the Aliev-Panfilov (right panels)
models; time is normalized by the action-potential duration (APD), which is
$5.92$ time units for the Panfilov model and $42.12$ time units for the
Aliev-Panfilov model. The upper panels show the initial transients in the time
series; the lower panels show a section of these time series in the
statistically steady, spiral-wave-turbulence states. The time series of $V$
consists of a train of action potentials. Note that the action potential for
the Aliev-Panfilov model does not have an overshoot region (Fig.~\ref{ap}(b)),
where the system repolarizes beyond its resting state.

%%%%%%%%%%%%%%%%%%%%%%%%%%%%%%%%%%%%%%%%%%%%%%%%%

Figure~\ref{2dpdf_vg} contains plots of PDFs of $V$ for (a) Panfilov and (b)
Aliev-Panfilov models and of $g$ for (c) Panfilov and (d) Aliev-Panfilov models
in 2D. Here and henceforth, in plots of PDFs, blue curves show PDFs and red
symbols the error bars. The PDF of $V$ for the Panfilov model shows two peaks
in the small-$V$ region. One peak corresponds to the overshoot region in the
time series and the other peak corresponds to the resting state. The peak at
high values of $V$ (or $g$) corresponds to the excited and the plateau regions.
These PDFs for the Aliev-Panfilov model have only one peak, in the small-$V$
region, at zero. This is because the Aliev-Panfilov model does not have an
overshoot region in its action potential; one peak in the PDF of $V$
corresponds to the resting state and the other peak, at a high value,
corresponds to the excited state and the plateau states. Error bars are small
and are barely visible on the scales of these figures. Note that these PDFs
depend on the details of these models; and they are markedly different from the
approximately Gaussian PDFs of velocity components in fluid
turbulence~\cite{pramanareview}. 

%%%%%%%%%%%%%%%%%%%%%%%%%%%%%%%%%%%%%%%%%%%%%%%%%%%%%%%%%%%%%%%%%%%%

Figure~\ref{2dpdf_gradvandg} contains semilogarithmic (base 10) plots of PDFs
of $(\nabla V)_x$ for (a) Panfilov and (b) Aliev-Panfilov models in 2D and PDFs
of $(\nabla g)_x$ for (c) Panfilov and (d) Aliev-Panfilov models in 2D.
Figure~\ref{2dpdf_gradvandg} shows that the PDFs of $\nabla V_x$ exhibit sharp
peaks at zero; the PDFs of $\nabla g_x$ have sharp
peaks at zero, like those in the PDFs of $\nabla V_x$; the PDFs of $(\nabla V)_y$ and $(\nabla g)_y$ for the Panfilov and the Aliev-Panfilov models are similar. These
gradient PDFs are sharply peaked at zero, indicating that, on average, these
gradients are small; however, these gradients have flat, non-Gaussian tails
and, in this regard, are qualitatively similar to the non-Gaussian tails of
PDFs of velocity gradients in fluid turbulence.

%%%%%%%%%%%%%%%%%%%%%%%%%%%%%%%%%%%%%%%%%%%%%%%%%%%%%%%%%%%%%%%%%

Figure~\ref{2dpdf_modvg} contains semilogarithmic (base 10) plots of the PDFs
of (a) $|\nabla V|$, (c) $| \nabla g|$ and (e) $\cos\theta = \frac{\nabla V .
\nabla g}{|\nabla V | |\nabla g |}$ for the 2D Panfilov model;
Figures~\ref{2dpdf_modvg} (b), (d) and (f) show, respectively, the
corresponding PDFs for the 2D Aliev-Panfilov model. For both these models, the
PDFs of $|\nabla V|$ and $|\nabla g|$ show dominant peaks at zero.
Figures~\ref{2dpdf_modvg} (e) and (f) show that, on average, $\nabla V$ and
$\nabla g$ tend either to align or to antialign with each other; the tendency
for being antialigned ($\cos\theta =-1$) is stronger than the tendency for
alignment ($\cos\theta = 1$). For cardiac models this may be a consequence of
the recovery processes (described by $g$) normally building up when the
excitation (variable $V$) has already decayed (see Fig.~\ref{2dts}). The
nearest analogs of these in fluid turbulence are PDFs of the angles between the
velocity and the vorticity in 3D fluid turbulence (see, e.g., Fig.~2d in
Ref.~\cite{ganapatimhd}).

%%%%%%%%%%%%%%%%%%%%%

Figure~\ref{2d_enerspec} contains log-log (base 10) plots of the spectra
$E_V(k)$ and $E_g(k)$ versus $k\lambda $; here $\lambda=10$, in dimensionless
units for the Panfilov model, and $ \lambda=48.5$, in dimensionless units for
the Aliev-Panfilov model, represents the wavelength of a plane wave in the
medium.  The dashed, black lines show (possible) power-law regime, with
$E_v(k)\sim k^{m_v} $ and $E_g(k) \sim k^{m_g} $ and $m_v \simeq -2.7 $
(Panfilov), $m_v \simeq -3.2$ (Aliev-Panfilov), $m_g \simeq -1.9$ (Panfilov),
$m_g \simeq -3.1$ (Aliev-Panfilov). The shapes of these spectra are
qualitatively similar in both models. In particular, we see two prominent peaks
in these spectra, located at $k\lambda = 1.4$ and $k\lambda = 1.7$ (Panfilov
model) and at $k\lambda = 2.1 $ and $ k\lambda = 2.4 $ (Aliev-Panfilov model).
However, these power-law regimes are certainly not as universal as their
fluid-turbulence counterparts in $E(k)$; nevertheless, the spectral slopes,
$-2.7$ and $-3.2$ for $E_V(k)$ are reasonably close to each other; thus, the
issue of their universality requires detailed, high-resolution studies in other
models and conditions. If these spectral exponents are not universal, we
conjecture that this is because of the large number of parameters in the models
we consider; in particular, this means that we cannot develop, in any obvious
way, the spiral-turbulence analog of the K41 phenomenological theory for the
scaling of fluid-turbulence, inertial-range energy spectra.

We have calculated the integral length scale $L_I$ and the Taylor-micro length $L_{\lambda}$ by using the spectrum $E_V(k)$. They are $2.62$ and $0.35$, respectively, for the 2D Panfilov model and $25.88$ and $0.39$, respectively, for the 2D Aliev-Panfilov model. By using the spectrum $E_g(k)$, we find $L_I$ and $L_{\lambda}$ to be $97.03$ and $0.50$, respectively, for the Panfilov model and $39.63$ and $0.43$, respectively, for the Aliev-Panfilov model. The inversus of these lengthscales are shown, by dashed lines, in the spectra given in Fig.~\ref{2dspectralengthscales}.

%%%%%%%%%%%%%%%%%%%%%%%%%%%%%%%%%%%%%%%%%%%%%%%%%%%%%%%%%%%%%%%%%%%%%%%%%%%%%%%%
 
In studies of fluid turbulence, it has been found that the power-law regime in
$S_p(r)$ can be \textit{extended} by using the \textit{extended self
similarity} (ESS) procedure, in which we plot $S_p(r)$ versus some other
structure function $S_{p1}(r)$. In Fig.~\ref{2dstruct_ext} we give such log-log
(base 10) extended-self-similarity (ESS) plots of $S_p(r)$ versus $S_2(r)$ of
$V$ for 2D (a) Panfilov and (c) Aliev-Panfilov models with $1 \leq p \leq 8$;
the dashed-black lines represent (possible) power-low regime. The power-law
ranges in these plots yield the multiscaling exponent ratios $\zeta_p/\zeta_2$,
which we plot versus $p$ in Fig.~\ref{2dstruct_ext} (b) and (d). In
Fig.~\ref{2dstruct_extg} we give such log-log (base 10)
extended-self-similarity (ESS) plots of $S_p(r)$ versus $S_2(r)$ of $g$ for 2D
(a) Panfilov and (c) Aliev-Panfilov models with $1 \leq p \leq 8$; the
dashed-black lines represent (possible) power-low regimes. The power-law ranges
in these plots yield the multiscaling exponent ratios $\zeta_p/\zeta_2$, which
we plot versus $p$ in Fig.~\ref{2dstruct_extg} (b) and (d). These exponent
ratios should be viewed with great caution, for they contain ratios of small
numbers; this becomes apparent if we plot structure functions versus $r$ and do
not use the ESS procedure. Whether or not such structure functions display
multiscaling can be settled only by DNSs that use much larger domains and
longer runs than those we have been able to perform. As we have mentioned in
the context of spectral exponents, the exponents or exponent ratios for the
Panfilov and Aliev-Panfilov models are comparable to each other, but not equal
at the resolution of our DNSs; the issue of their universality or lack thereof
requires detailed, high-resolution studies in other models and conditions. 

%%%%%%%%%%%%%%%%%%%%%%%%%%%%%%%%%%%%%%%%%%%%%%%%%%%%%%%%%%%%%%%%%%%%%%%%

\subsection{Three-dimensional results for scroll-wave turbulence}

Our DNSs in 3D use a domain of size $640 \times 640 \times 640 $. We begin with
half the medium excited and the other half in the excitable state; and then we
apply a stimulus at the middle of the domain to form a scroll wave, which we
use as an initial condition (Fig.~\ref{3dsnap_c}). Figure~\ref{3dsnap_p}
presents pseudocolor plots of the transmembrane potential $V$ showing the
scroll waves we use as initial conditions for our simulations of the 3D
Panfilov (top, left panel) and the Aliev-Panfilov (bottom, left panel) models
and typical turbulent states in our simulations of the 3D Panfilov (top, right
panel) and Aliev-Panfilov (bottom, right panel) models, at a representative
time, 3.8 (dimensionless units), in the turbulent, statistically steady state,
which is, to a good approximation, homogeneous and isotropic far away from
boundaries. Given our set of parameter values, the scroll-wave initial
conditions have evolved into small, broken scrolls that interact with each
other. After the system has reached a statistically steady state, we record the
values of $V$ and $g$ and calculate the statistical properties of this state.
We present our results in Figs.~\ref{3dts}-\ref{3dstruct_extg}.

In Fig.~\ref{3dsnap_c} we portray the initial and final states of our 3D
system for the two models by using isosurface plots of $V$ for
$V=0.8$ (dimensionless units). Note that the scroll-wave initial conditions have
evolved into small, broken scrolls that interact with each other. This is the
analog of the Richardson cascade in fluid turbulence. Instead of large eddies
breaking down into ever smaller eddies, here a large scroll wave breaks down
into smaller and smaller scroll waves. These isosurfaces of $V$ should be
contrasted with the tubular structures of isosurfaces of the modulus of the
vorticity in 3D fluid turbulence~\cite{kaneda,pramanareview,ganapatimhd}.

%%%%%%%%%%%%%%%%%%%%%%%%%%%%%%%%%%%%%%%%%%%%%%%%%

Figure~\ref{3dts} shows local time series of $V$ (blue lines) and $g$ (red
dashed lines), obtained from the representative point $(x=200, y=200, z=300)$
(in dimensionless units) from our 3D domains for the Panfilov (left panels) and
the Aliev-Panfilov (right panels) models; time is normalized by the
action-potential duration (APD), which is $5.92$ time units, for the Panfilov
model, and $42.12$ time units, for the Aliev-Panfilov model. The upper panels
show the initial transients in these time series; the lower panels show a
section of these time series in the statistically steady,
scroll-wave-turbulence states. The time series of $V$ consists of a train of
action potentials. The action potential for the Aliev-Panfilov model does not
have an overshoot region (see Fig.~\ref{ap}), where the system repolarizes
beyond its resting state.

Figure~\ref{3dpdf_vg} contains plots of PDFs of $V$ for (a) Panfilov and (b)
Aliev-Panfilov models and of $g$ for (c) Panfilov and (d) Aliev-Panfilov models
in 3D. As in our 2D simulation, the PDF of $V$, for the Panfilov model in 3D,
shows two peaks in the small-$V$ region. One peak corresponds to the overshoot
region in the APD and the other corresponds to the resting state. The peak at
a high value of $V$ (or $g$) corresponds to the excited and the plateau region.
These PDFs for the Aliev-Panfilov model have a peak in the small-$V$ region at
$V=0$. The Aliev-Panfilov model does not have an overshoot region in its action
potential. Its one peak corresponds to the resting state and the other peak, at
a high value of $V$, corresponds to the excited and the plateau states.

Figure~\ref{3dpdf_gradvandg} contains semilogarithmic (base-10) plots of PDFs
of $(\nabla V)_x$ for (a) Panfilov and (b) Aliev-Panfilov models in 3D. These
figures show dominant peaks at zero. The PDFs of $(\nabla V)_y$ and of $(\nabla
V)_z$ are similar. In Fig.~\ref{3dpdf_gradvandg} the bottom panels contain
semilogarithmic (base 10) plots of PDFs of $(\nabla g)_x$ for (c) Panfilov and
(d) Aliev-Panfilov models in 3D. The dominant peaks are at zero. The PDFs of
$(\nabla g)_y$ and $(\nabla g)_z$ are similar.

Figure~\ref{3dpdf_modvg} contains plots of the PDFs of (a)$|\nabla V|$, (c)$|
\nabla g|$, and (e) $\cos\theta = \frac{\nabla V \cdot \nabla g}{|\nabla V |
|\nabla g |}$ for the 3D Panfilov model; (b), (d), and (f) show, respectively,
the corresponding PDFs for the 3D Aliev-Panfilov model. For both these models,
the PDFs of $|\nabla V|$ and $|\nabla g|$ show dominant peaks at zero.
Figures~\ref{3dpdf_modvg} (e) and (f) show that, on average, $\nabla V$ and
$\nabla g$ tend either to align or to anti-align with each other. The degree of
alignment or anti-alignment is different in these two models; for the Panfilov
model, the tendency for being anti-aligned ($\cos\theta =-1$) is stronger than
the tendency for alignment ($\cos\theta = 1$).

Figure~\ref{3d_enerspec} contains log-log (base 10) plots of the spectra
$E_V(k)$ and $E_g(k)$ versus $k\lambda $; here $\lambda=10$, in dimensionless
units for the Panfilov model, and $ \lambda=48.5$, in dimensionless units for
the Aliev-Panfilov model. 
%represents the wavelength of a plane wave in the medium. 
The dashed, black line shows a (possible) power-law regimes, with
$E_v(k)\sim k^{m_v} $ and $E_g(k) \sim k^{m_g} $ and $m_v \simeq -3.8$
 (Panfilov), $m_v \simeq -3.9$ (Aliev-Panfilov), $m_g \simeq -1.8$ (Panfilov),
and $m_g \simeq -4.4$ (Aliev-Panfilov). The shapes of these spectra are
qualitatively similar in both models; in particular, we see one sharp feature 
in these spectra, located at $k\lambda = 1.7$ (Panfilov model) and at $
k\lambda = 2.4 $ (Aliev-Panfilov model). The spectral slopes, $-3.8$ and
$-3.9$, for $E_V(k)$ are reasonably close to each other; but, as we have noted
in our discussion of our 2D-spiral-wave-turbulence results, the issue of the
universality of these exponents requires detailed, high-resolution studies in
other models and conditions.

We have calculated the integral length scale $L_I$ and the Taylor-micro length $L_{\lambda}$ by using the spectrum $E_V(k)$. They are $3.28$ and $1.39$, respectively, for the 3D Panfilov model and $4.95$ and $2.41$, respectively, for the 3D Aliev-Panfilov model. By using the spectrum $E_g(k)$, we find $L_I$ and $L_{\lambda}$ to be $2.01$ and $0.52$, respectively, for the Panfilov model and $4.26$ and $1.98$, respectively, for the Aliev-Panfilov model. The inversus of these lengthscales are shown, by dashed lines, in the spectra given in Fig.~\ref{3dspectralengthscales}.

In Fig.~\ref{3dstruct_ext} we give log-log (base 10) extended-self-similarity
(ESS) plots of $S_p(r)$ versus $S_2(r)$ of $V$ for 3D (a) Panfilov and (c)
Aliev-Panfilov models with $1 \leq p \leq 8$. The dashed-black lines represent
(possible) power-low regimes. The power-law ranges in these plots yield the
multiscaling exponent ratios $\zeta_p/\zeta_2$, which we plot versus $p$ in
Fig.~\ref{3dstruct_ext} (b) and (d). In Fig.~\ref{3dstruct_extg} we give such
log-log (base 10) extended-self-similarity (ESS) plots of $S_p(r)$ versus
$S_2(r)$ of $g$ for 3D (a) Panfilov and (c) Aliev-Panfilov models with $1 \leq
p \leq 8$. The power-law ranges in these plots yield the multiscaling exponent
ratios $\zeta_p/\zeta_2$, which we plot versus $p$ in Fig.~\ref{3dstruct_extg}
(b) and (d). Our cautionary remarks, for ESS plots in the case of 2D, 
spiral-wave turbulence, apply here too.

%%%%%%%%%%%%%%

\subsection{Scroll-wave turbulence with negative filament tension}

\begin{figure}
	\includegraphics[width=1\linewidth]{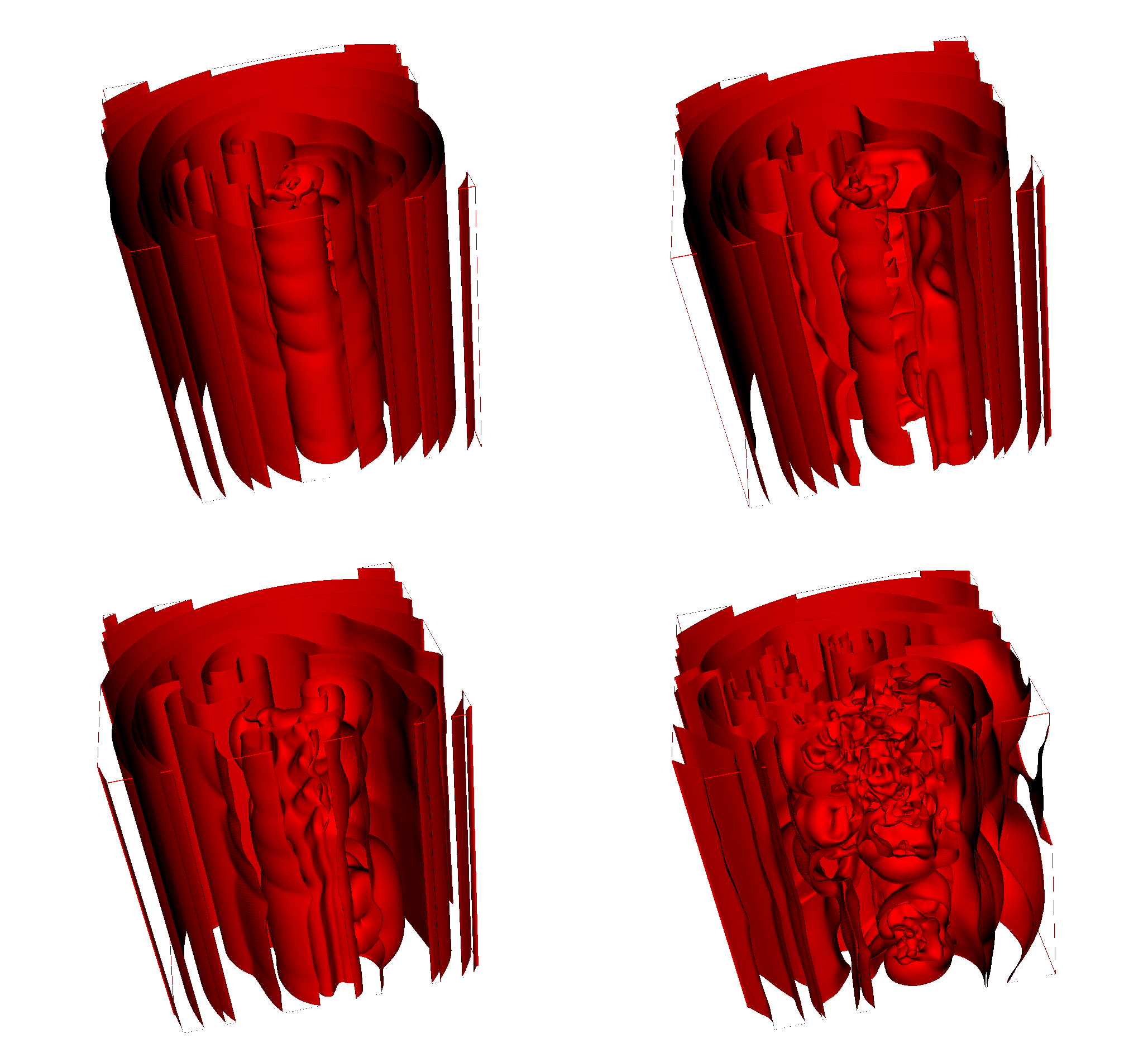}
\caption{Isosurface plots of scroll-wave formation and subsequent break up in
the case of the Aliev-Panfilov model with parameters such that the scroll
filament has a negative tension, which is manifest here in the transverse
undulations in the scroll sheets. For the complete spatio temporal evolution of th scrolls, see movie S9 in the Supplementary Material at ~\cite{SuppMat}.} \label{negfil} 
\end{figure}
\begin{figure*}
\includegraphics[width=1\linewidth]{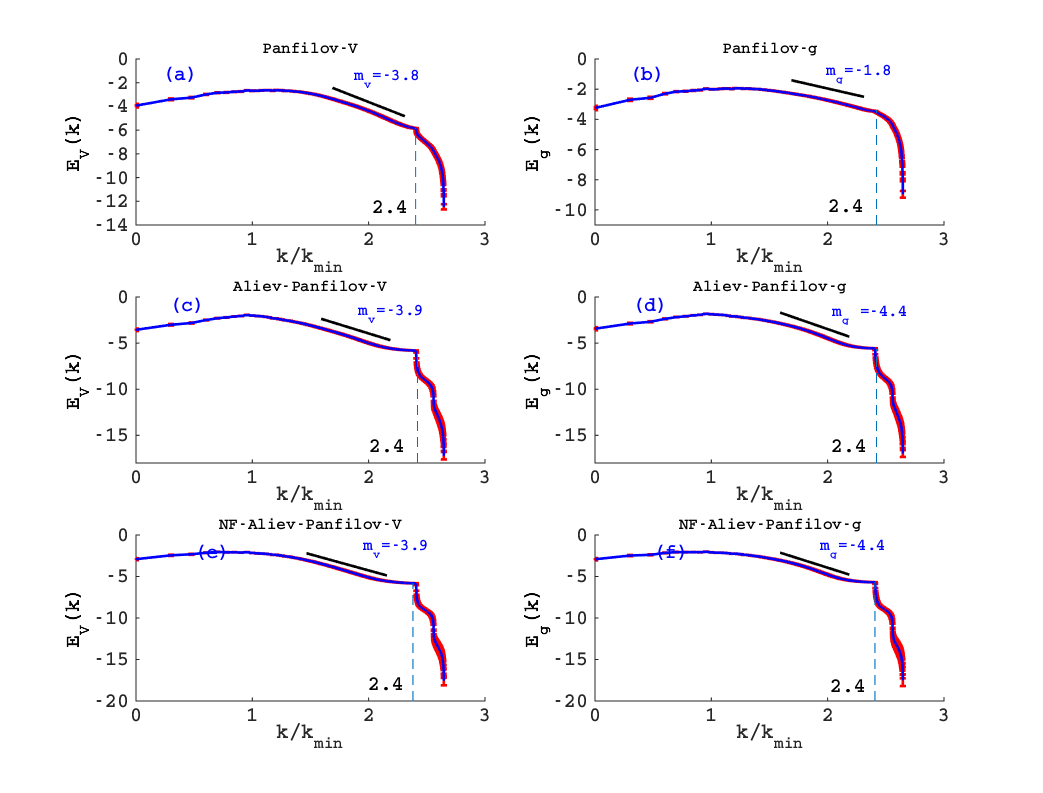}
\caption{Comparison of the spectra of $E_V$ and $E_g$ for 3D-scroll-wave
turbulence in the three models
that we have studied. Here we show $k/k_{min}$ on the horizontal axes;
the peaks of the spectra occur at the same value of $k$, namely, 
$k/k_{min} \simeq 2.4$.}
\label{spec-neg}
\end{figure*}
\begin{figure}
\begin{center}
\includegraphics[width=1\linewidth]{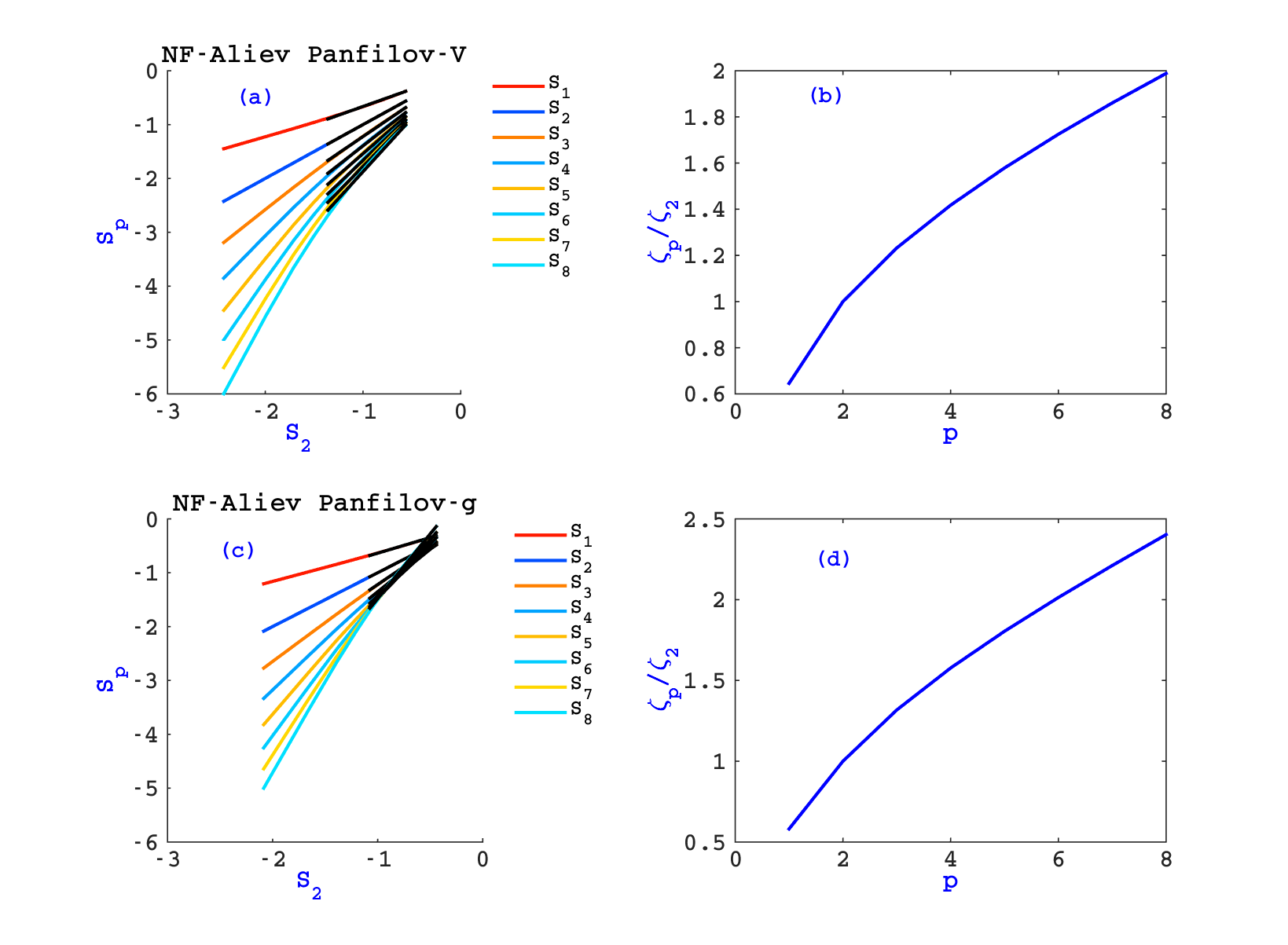}
\caption{ESS plots of structure functions of $V$ and $g$ 
for the case of 3D-scroll-wave turbulence in the 
negative-filament-tension regime.}
\label{ess-neg}
\end{center}
\end{figure}
%%%%%%%%%%%%%%%%%%%%%%%%%%%%%

The breakdown of scroll waves has been studied in a model, for cardiac tissue,
that displays negative filament tension~\cite{Alonso08}. Such studies have not
used simulation domains that are large enough to yield statistically
homogeneous and isotropic scroll-wave turbulence. Therefore, we have carried
out a study of scroll-wave turbulence in a parameter regime in the
Aliev-Panfilov model in which the filament tension is negative. In this case,
the scroll sheets show undulations in the transverse direction, which indicates
that the scroll filament has a negative tension; it first becomes elongated,
and then it breaks up; this is shown in Fig.~\ref{negfil}. The parameter set
for the Aliev-Panfilov model, which gives rise to this behavior of the
scroll-waves, is $\mu_1 = 0.05; \mu_2 = 0.30;$ and $g=g_{min}*74$, where
$g_{min}=0.033$. We have calculated the spectra $E_V$ and $E_g$ for this case
and found that, as far as statistical properties of scroll-wave turbulence are
concerned, there is no significant difference between the statistical
properties of 3D-scroll-wave turbulence, with and without negative tension for
the scroll filament. This interesting universality of scroll-wave turbulence
(i.e., spectral exponents that do not seem to depend on the filament tension of
the initial scroll wave) is illustrated in the log-log plots of
Fig.~\ref{spec-neg}, which compare the spectra  $E_V$ and $E_g$ for scroll-wave
turbulence with and without negative filament tension.  Figure~\ref{ess-neg}
shows a comparison of ESS plots of structure functions for both these cases.

%???IS THAT ALSO log log plot as in Fig.19, if so we need to mention it in the
%legend. ALSO Neg and positive tension look almost identical. I think this is
%very interesting. We can say that turbulence is universal independently on the
%mechanism, important that we have may wavelets here, but not the way how these
%wavelets were formed. MAYBE we add something like that to the text???

%%%%%%%%%%%%%%%%%%%%%%%%%%%%%%%%%%%
\section{Conclusion}

We have carried out the most extensive numerical study, attempted so far, of
the statistical properties of scroll-and spiral- wave turbulence in two simple,
two-variable models for cardiac tissue. In particular, we have examined, via
extensive direct numerical simulations (DNSs), the statistical properties of
spiral- and scroll-wave turbulence in two- and three-dimensional excitable
media by using the Panfilov~\cite{Panfilov-model} and the
Aliev-Panfilov~\cite{Aliev-Panfilov-model} mathematical models for cardiac
tissue. We use very large simulation domains, made possible by our
state-of-the-art simulations on graphics processing units
(GPUs)~\cite{cuda-man}, with a view to comparing the statistical properties of
spiral- and scroll-wave turbulence here with the statistical properties of
statistically homogeneous and isotropic two- and three-dimensional fluid
turbulence. We hope our study of the statistical properties of homogeneous and
isotropic spiral- and scroll-wave turbulence will stimulate experimental
studies of such turbulence.  

As we have noted above, there is an important qualitative difference between
spiral- and scroll-wave turbulence in excitable media and fluid turbulence
insofar as turbulence in an excitable medium is neither decaying nor forced.
Once turbulence has been initiated in an excitable medium, e.g., by a large
spiral or scroll wave, we have shown that there is a forward cascade, which
yields small spirals or scrolls; these form, interact, and break all the time.
We have shown that the resulting turbulent state is statistically steady and,
far away from boundaries, is statistically homogeneous and isotropic. For the
fields $V$ and $g$, we have shown that the spectral analogs of $E(k)$ in fluid
turbulence, are spread out over several decades in $k$. Therefore, we have
demonstrated that, like fluid turbulence, spiral- and scroll-wave turbulence
involves a wide range of spatial scales; $E_V(k)$ and $E_g(k)$ show approximate
power laws in some range of $k$, but the exponents for these do not appear to
be as universal as their counterparts in fluid turbulence. Even though there
are diffusive terms in the equations we consider, they do not dissipate spiral
or scroll waves completely because of the excitability of the medium. No
external forcing is required to maintain these states of spiral- or scroll-wave
turbulence. The only requirements are (i) a suitable initial condition and (ii)
$L/\lambda \to \infty$ (or $L/\lambda$ large enough in a practical
calculation), where $L$ is the linear size of the simulation domain and
$\lambda$ the wavelength of a plane wave in the excitable medium; the
dimensionless ratio $L/\lambda$ is a convenient, Reynolds-number-type control
parameter ($L$ is also used as a control parameter in a suitably scaled version
of the Kuramoto-Sivashinsky equation~\cite{jay93}). 

{{As we have mentioned above, 2D and 3D fluid turbulence are qualitatively different in so far as
the former displays an inverse cascade of energy whereas the latter shows a forward cascade of
energy. Spiral- and scroll-wave turbulence are not different in this manner. In both these cases there is the analog of a forward cascade: in 2D, large spirals break into small ones and in 3D, large scrolls also break into small ones. However, there are important differences between  the spectral properties and PDFs that characterize the statistical properties of 2D, spiral-wave turbulence and 3D, scroll-wave turbulence. The differences can be gleaned by comparing our figures in Sec.III.A-B with their counterparts in Sec.III.C.
}}

\section{Acknowledgments}

We thank R. Majumder and A.R. Nayak for discussions, Council of Scientific and
Industrial Research, University Grants Commission, Department of Science and
Technology, and the European Cooperation in Science and Technology Action MP006
for support, and Supercomputing Education and Research Centre (IISc) for
computational resources. 

%%%%#####################################################################################

\end{document}